\documentclass[preprint,3p,twocolumn]{elsarticle}

\usepackage{lineno,hyperref}
\modulolinenumbers[1]
\usepackage{amsmath}
\usepackage{siunitx}
\biboptions{numbers,sort&compress}

\usepackage{tikz} %for drawing the energy flow diagram
\usepackage[europeanresistors,americaninductors]{circuitikz}
\usepackage{adjustbox}
\usepackage{eurosym}
\usepackage{xspace}
\usepackage{caption}
\usepackage{booktabs}
\usepackage{tabularx}
\usepackage{threeparttable}
\usepackage{multicol}
\usepackage{float}
\usepackage{graphicx,dblfloatfix}
\usepackage{csvsimple}
\usepackage{url}
\usepackage{lipsum}

\graphicspath{{figures/}}

\journal{Applied Energy}

\newcommand{\ubar}[1]{\text{\b{$#1$}}}

\newcommand{\WH}{\text{weakly homogeneous}\xspace}

\def\co{CO${}_2$}
\def\el{${}_{\textrm{el}}$}
\def\th{${}_{\textrm{th}}$}

\begin{document}

\begin{frontmatter}

\title{Impact of \co{} prices on the design of a highly decarbonised \protect\\ 
	coupled electricity and heating system in Europe}

\author[aarhus]{K.~Zhu\corref{cor1}}
\ead{kunzhu@eng.au.dk}
\author[aarhus]{M.~Victoria}
\author[kit]{T.~Brown}
\author[aarhus]{G. B.~Andresen}
\author[aarhus]{M.~Greiner}

\cortext[cor1]{Corresponding author}

\address[aarhus]{Department of Engineering, Aarhus University, 8000 Aarhus C, Denmark}
\address[kit]{Institute for Automation and Applied Informatics (IAI), Karlsruhe Institute of Technology (KIT), Forschungszentrum 449,76344, Eggenstein-Leopoldshafen, Germany}

\begin{abstract}
Ambitious targets for renewable energy and \co{} taxation both represent political instruments for decarbonisation of the energy system. We model a high number of coupled electricity and heating systems, where the primary sources of \co{} neutral energy are from variable renewable energy sources (VRES), \textit{i.e.}, wind and solar generators. The model includes hourly dispatch of all technologies for a full year for every country in Europe. In each model run, the amount of renewable energy and the level of \co{} tax are fixed exogenously, while the cost-optimal composition of energy generation, conversion, transmission and storage technologies and the corresponding \co{} emissions are calculated. We show that even for high penetrations of VRES, a significant \co{} tax of more than 100~\euro/t\co{} is required to limit the combined \co{} emissions from the sectors to less than 5\% of 1990 levels, because curtailment of VRES, combustion of fossil fuels and inefficient conversion technologies are economically favoured despite the presence of abundant VRES. A sufficiently high \co{} tax results in the more efficient use of VRES by means of heat pumps and hot water storage, in particular. We conclude that a renewable energy target on its own is not sufficient; in addition, a \co{} tax is required to decarbonise the electricity and heating sectors and incentivise the least cost combination of flexible and efficient energy conversion and storage. 
\end{abstract}


\end{frontmatter}

%\linenumbers

\section{Introduction}
In order to limit the increase of the global average temperature to \SI{2}{\celsius} it is mandatory to reduce, drastically and in a short time, anthropogenic \co{} emissions. The cost reduction of renewable energy sources during the last decade, particularly that of wind and photovoltaics (PV), has paved the way to decarbonise the generation of electricity. %An extensive literature exists \cite{Schaber_2012,Schaber_2012b,Scholz_thesis,Rasmussen2012,Haller_2012,Rodriguez_2014,Bussar_2014,Eriksen2017,Schlachtberger_2017,Gils2017a,Cebulla_2017,Breyer_2018,Collins_2018,Brown_2018b,schlachtberger2018cost} devoted to analyse a highly renewable European power system. Some of the authors of this work have previously dealt with this problem, investigating the impact of different renewable capacity layouts \cite{Eriksen2017}, as well as the role of storage and interconnections to counterbalance the inherent variability of wind and PV generation \cite{Rasmussen2012,Rodriguez_2014,Schlachtberger_2017,schlachtberger2018cost}.

However, drastically reducing greenhouse gas emissions in the power system will not be enough to avoid exceeding the \SI{2}{\celsius} limit. A fast \co{} reduction in other sectors is also required \cite{Rogelj_2015}. This represents an opportunity to exploit the synergies of sector coupling \cite{Brown_2018, Wu_2017, Lund_2017}. In this study, we investigate the decarbonisation of coupled electricity and heating sectors in Europe. The annual end use demands in both sectors are similar: electricity consumption in Europe in 2015 accounted for 2,854 TWh\el (in terms of electricity), while the heating demand in the residential and services sectors represented 3,562 TWh\th (in terms of thermal energy), \cite{HRE}. Regarding greenhouse gas emissions, electricity and low-temperature heating accounted for 1066 Mt and 556 Mt of \co{} emissions respectively \cite{EEA}.

Previous energy models applied to different regions \cite{Brown_2018, Wu_2017, Jacobson_2011, Connolly2016, Ashfaq_2017} have shown that imposing a strong \co{} constraint leads to high Variable Renewable Energy Sources (VRES) penetrations combined with a high-efficiency sector-coupled energy system, referred to as `Smart Energy System' by Lund and coauthors \cite{Connolly2016, Lund_2017}. For instance, under a 95\% \co{} reduction constraint relative to 1990 in Europe \cite{roadmap2050}, energy modelling approaches based on scenario comparison \cite{Connolly2016} or cost optimisation \cite{Brown_2018} reach similar configurations, \textit{e.g.}, heat pumps converting electricity into heat, and district heating systems fed with Combined Heat and Power (CHP) plants are the key-enabling technologies to decarbonise the heating sector efficiently. In particular with regard to the decarbonisation of the heating sector, some authors have analysed the possibility of using excess renewable electricity to provide heat \cite{Meibom_2007,Pensini_2014} but no constraint on the \co{} emission from this sector was imposed. In \cite{Ashfaq_2017}, the coupled electricity-heating system has been investigated using a simplified investment and dispatch algorithm for regional, country and pan-European networks. Coupling electricity and heating sectors might provide additional advantages. Prominent among them is the fact that thermal energy storage adds flexibility to the system at a very low cost, which may reduce the requirements of increasing interconnections among adjacent countries.

It should be mentioned that alternative scenarios compatible with the \SI{2}{\celsius} target, and assuming a more modest contribution from VRES, have been proposed for Europe \cite{TIMES} and at a global scale \cite{IPCC_Energy, Rockstrom_2017, Bertram_2015}. In those scenarios, a significant contribution is expected from biomass and fossil fuel-fired power plants including Carbon Capture and Storage (CCS) and, in some cases, from nuclear power plants. The former entails notable uncertainties associated with CCS technological feasibility and future cost projections. In this study, we focus on a sector-coupled system based on VRES, that is, wind and solar. By using a simplified cross-sector network model, we are able to capture the more general system dynamics and extract meaningful insights, as well as inspirational results, that can be further investigated in subsequent works. 

Governments have implemented several economic instruments to incentivise the different stakeholders so that the \co{} reduction target can be reached in due time. In the EU's 2020 package, a 20\% target for renewable energy was set to lead the initial steps in the energy transition for most European countries. Nevertheless, some authors argue that it is economically more efficient to price \co{} emissions \cite{Jonghe_2009}. In reality, the EU Emission Trading System (ETS) implemented in 2005 has so far not been fully effective due to the low \co{} market prices. ETS is also not effective since it does not cover all sectors, \textit{e.g.}, residential heating or transportation. Alternative instruments proposed to incentivise \co{} reductions aim to force a more stable \co{} price through taxation \cite{Bertram_2015, Pezzey_2013, Wilson_2018}. The benefits of combining renewable energy targets while simultaneously pricing \co{} emissions, have also been highlighted by some researchers \cite{Bertram_2015, Rio_2017}. A practical implementation of these instruments with a very positive outcome can be found in Great Britain where the introduction of a Carbon Price Floor (CPF) has allowed a rapid fuel switching from coal to natural gas, halving the power sector emissions in just 4 years \cite{Wilson_2018}. The Carbon Price Support policy introduced in 2013, stabilises the \co{} price by requiring power-sector emitters to pay a top-up price to a CPF determined by policy makers and whose value is currently set at 18~\pounds/t\co{} (20.3~\euro/t\co{}). The summary of carbon pricing initiatives published by the World Bank includes carbon tax values from different countries ranging up to 140~\$/t\co{} (123~\euro/t\co{}) in the case of Sweden \cite{Carbon_pricing_WB}. Although \co{} price in the order of tens of euros per ton could be enough to disincentivise the use of coal in the power sector, the optimum price may be totally different when considering the coupled sectors and a strong \co{} reduction target. In \cite{Brown_2018} a shadow price of 407~\euro/t\co{} was estimated when optimising the coupled electricity, heating, and transportation system under a 95\% \co{} reduction constraint. This result indicates that a \co{} price of this order of magnitude could act as an economic incentive to bring about the intended transition.

In this paper, we investigate the combined effect of a constrained VRES generation and \co{} price in the coupled electricity-heating system of Europe. The main research question can be formulated as follows: 
Is increasing the VRES penetration enough to achieve a low-\co{} system or are other measures such as \co{} pricing necessary?
This also allows us to answer a secondary and more specific question: 
If the VRES penetration is fixed, what is the required \co{} price to largely reduce the use of gas in the system and enable \SI{2}{\celsius}-compatible supply of the electricity and heating demands? 

\begin{figure}[!h]
	\centering
	\includegraphics[trim=0 0cm 0 0,width=\linewidth,clip=true]{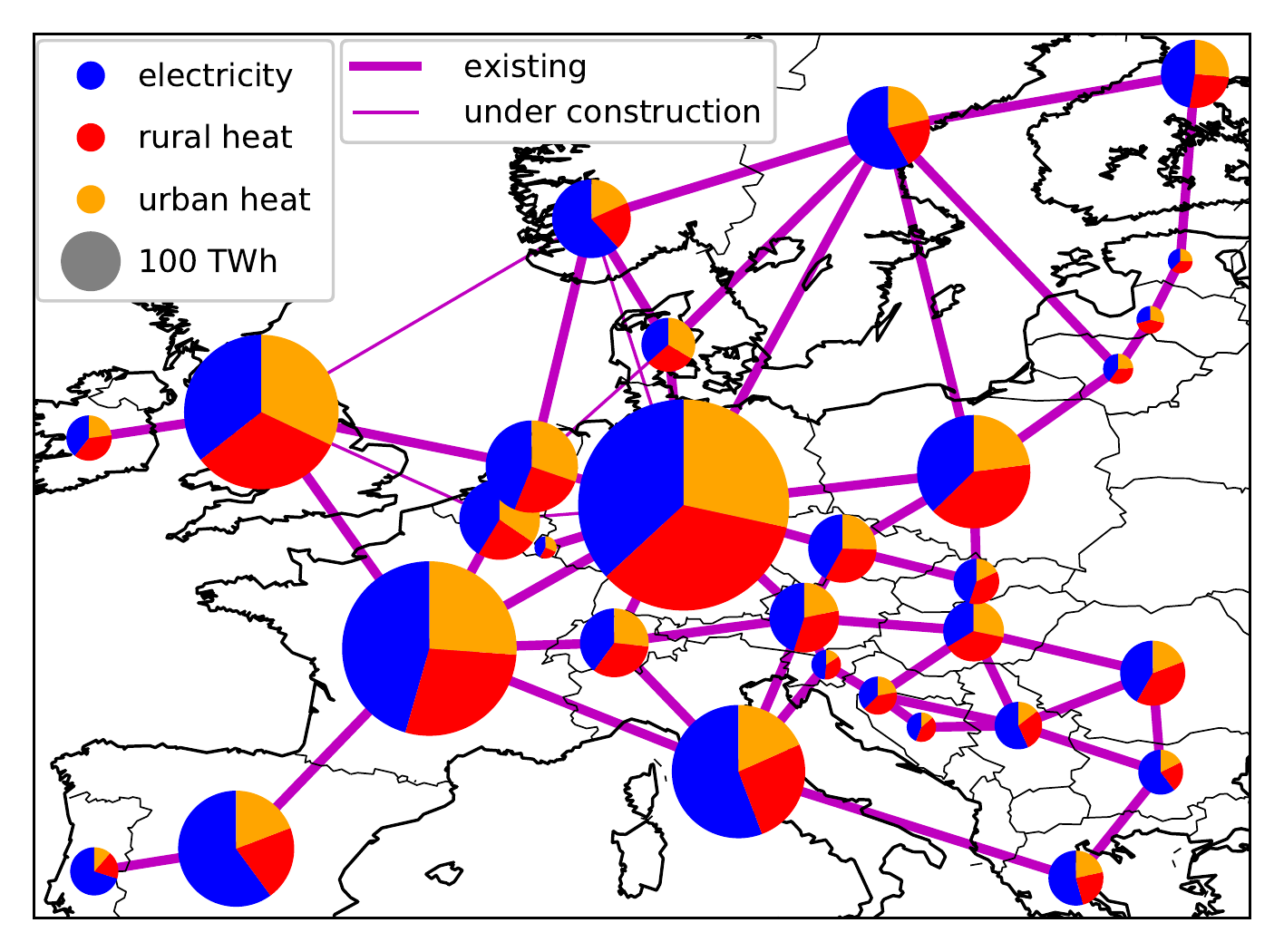}
	\includegraphics[trim=0 0cm 0 0,width=\linewidth,clip=true]{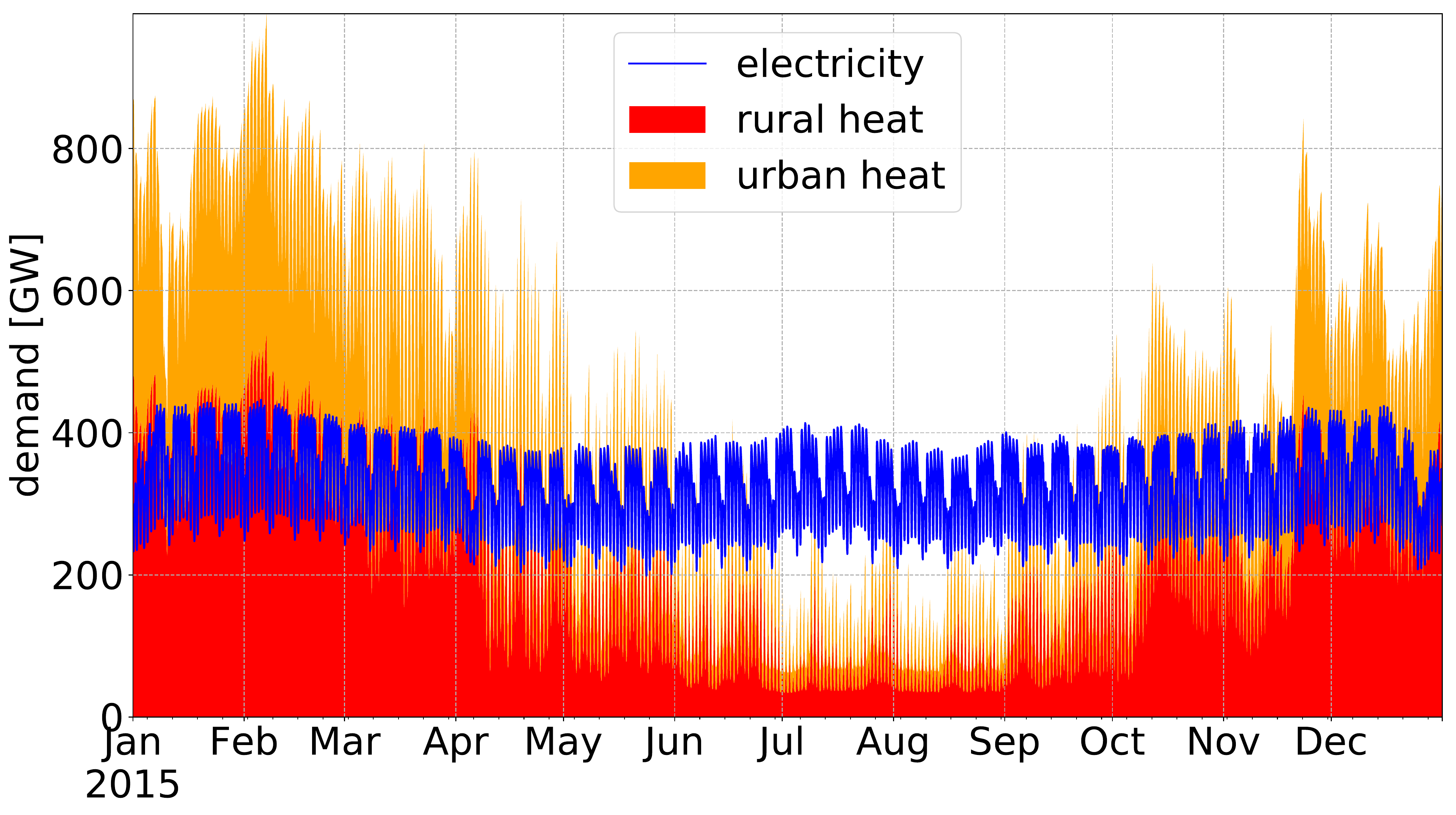}
	\caption{Annual national demands and transmission grid map (top) and aggregated European demand profiles (bottom). Demands consist of electricity (2854~TWh\el), heat in high population-density areas (urban heat, 1624~TWh\th), and heat in low population-density areas (rural heat, 1939~TWh\th). Transmission lines include existing and under construction lines (cross-border lines from ENTSO-e \cite{ENTSOE} grid map). Aggregated European electricity and heating demand profiles are from 2015. The electricity demand is historical data gathered from ENTSO-e, whereas heating demand is based on the heating-degree-hour approximation.}
	\label{fig:topology and demand}
\end{figure}

The major novelty of this paper resides in its global approach where the combined effects of interconnections among European countries, sector couplings, constraints on the VRES penetration, and \co{} prices are simultaneously investigated. The European model described in \cite{Brown_2018} is used. A network with 30 nodes, each of them representing a country in Europe, is assumed. The analysis of the cost-optimal system configuration obtained in \cite{Brown_2018} for the coupled electricity-and-heating system is extended here by performing extensive sweeps through parameters such as the renewable penetration, the wind/solar mix, and the \co{} price. Figure \ref{fig:topology and demand} shows the topology and transmission grid. A fundamental difference to \cite{Brown_2018} is that, the annual VRES generation in every country is proportional to the average demand of the country. In essence, it represents a plausible future situation, where, due to the limited interconnections, the European countries need to be relatively self-sufficient while decarbonising their energy systems. To the best of our knowledge, this is the first time that the analysis of the described problem including all the relevant elements is tackled enabling us to provide consistent answers to the research questions.

%This paper connects naturally to the extensive research on weather-driven modelling of power systems \cite{Eriksen2017, Rasmussen2012} carried out by some of the authors of this paper. In that case, a set of operational rules are implemented to ensure the most efficient integration of renewable energy providing an upper bound for their used and consequently, an upper bound on possible \co{} emissions. For power system, the operational rules are not very different from a marginal price-based market but the weather-based type of modelling becomes complicated in the context of sector coupled 'smart energy systems', where many alternative energy conversion and storage solutions exist. Often the more efficient solution also comes at a higher cost, e.g., a heat pump versus an electrical boiler, and the impacts of investment and operation costs are strongly linked. In the present work, we resolve this issue by introducing an extremely high tax on \co{} in some scenarios to identify the most efficient combined system. This can then be compared to designs with lower \co{} tax, which would typically be less efficient but cheaper.

The paper is organised as follows. Section \ref{sec:methods model} describes the model including the simulation framework utilised and the hypotheses on the renewable capacity configuration. Section \ref{sec:methods data} summarises the input data including costs, efficiencies, and other parameters assumed for the different technologies. Section \ref{sec:result} and Section \ref{sec:discussion} present the results and discussion. Finally, Section \ref{sec:conclusion} gathers together the main conclusions and opens for several possible future extensions.

%-------------------------------
\section{Methods: Model} \label{sec:methods model}
\begin{figure*}[!h]
%\captionsetup{labelformat=empty}
%\caption{Energy flow diagram of one node}
%trim={<left> <lower> <right> <upper>}
\centering
\begin{adjustbox}{scale=0.8,trim=0 9cm 0 0}
\begin{circuitikz}
%electricity bus
\draw (4,14.5) to [short,i^=import] (4,13);
\draw [ultra thick] (-7,13) node[anchor=south west]{\textbf{electricity}} -- (7,13);
\draw (5,13) |- +(0,0.5) to [short,i^=export] +(2,0.5);
\draw (0,-0.5) ;
\draw (-1,13) -- +(0,-0.5) node[sground]{};
\draw (0,14) node[vsourcesinshape, rotate=90](V2){}(V2.left) -- +(0,-0.6);
\draw (1.4,14.2) node{wind, solar;};
\draw (1.6,13.8) node{hydro, OCGT;};
\draw (1.3,13.4) node{CHP elec};
\node[draw,minimum width=1cm,minimum height=0.6cm,anchor=south west] at (-5,13.5){battery, hydro, hydrogen};
\draw (-2.9,13) to (-2.9,13.5);
%rural heat bus
\draw [ultra thick] (-7,10) node[anchor=south west]{\textbf{rural heat}} -- (-0.5,10) ;
\draw (-4,10) -- +(0,-0.5) node[sground]{};
%\draw (-3.5,9) node[vsourcesinshape, rotate=270](V2){}(V2.left) -- +(0,0.6);
\draw (-5,13) to [short,i^=ground heat pump;] (-5,10);
\draw (-3.7,11) node{resistive heater};
\node[draw,minimum width=1cm,minimum height=0.6cm,anchor=south west] at (-2.5,8.7){short-term storage};
\draw (-1,10) to (-1,9.3);
%urban heat bus
\draw [ultra thick] (0.5,10)  -- (7,10) node[anchor=south east]{\textbf{urban heat}};
\draw (2,10) -- +(0,-0.5) node[sground]{};
%\draw (3,9) node[vsourcesinshape, rotate=270](V2){}(V2.left) -- +(0,0.6);
\draw (4.5,13) to [short,i^=air heat pump;] (4.5,10);
\draw (6.1,12.1) node{ground heat pump;};
\draw (5.8,11) node{resistive heater};
\node[draw,minimum width=1cm,minimum height=0.6cm,anchor=south west] at (3.6,8.7){long-term storage};
\draw (5,10) to (5,9.3);
%conventionals are considered to be generators
\draw (1.5,11) node[vsourcesinshape, rotate=90](V2){}(V2.left) -- +(0,-0.6);
\draw (2.8,11.2) node{CHP heat};
\draw (2.8,10.8) node{gas boiler};
\draw (-1.5,11) node[vsourcesinshape, rotate=90](V2){}(V2.left) -- +(0,-0.6);
\draw (-0.2,11) node{gas boiler};
%\node[draw,minimum width=1cm,minimum height=0.6cm,anchor=south west] at (1,11.2){gas};
%\draw (1.5,11.8) to [short,i^=${}$] (1.5,13);
%\draw (2.5,12.5) node{CHP elec;};
%\draw (2.5,12.1) node{OCGT};
%\draw (1.5,11.2) to [short,i^=${}$] (1.5,10);
%\draw (2.5,10.8) node{CHP heat};
%\draw (2.5,10.4) node{gas boiler;};
%\draw (1.5,11) |- +(-3.5,0) to [short,i^=gas boiler] +(-3.5,-1);
\end{circuitikz}
\end{adjustbox}
\caption{Energy flow diagram of one node, representing a European country. Each node is divided into three buses: one electricity bus and two heat buses named rural and urban, according to population density. Loads (triangles), generators (circles), storage units (rectangles), and transmission lines and energy converters (lines with arrows) are attached to buses.} 
\label{fig:flow}
\end{figure*}
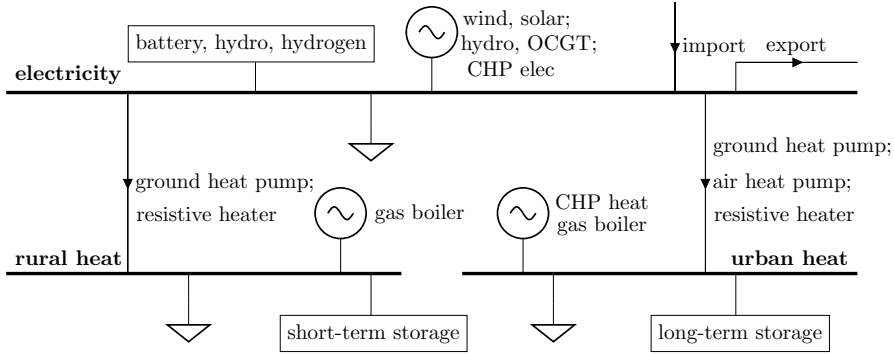

%-------------------------------
The model is implemented as a techno-economical optimisation problem, which minimises the total system costs expressed as a linear function (Eq.\ (\ref{eq:objective})) subject to technical and physical constraints (Eqs.\ (\ref{eq:energybalance}) - (\ref{eq:soc})), assuming perfect competition and foresight. The open-source framework PyPSA \cite{PyPSA} and the PyPSA-Eur-Sec-30 model introduced in \cite{Brown_2018}, are used. Due to computational limits, each of the 30 European countries covered by the model is aggregated into one node, which consists of one electricity and two heat buses. Neighbouring countries are connected through cross-border transmission lines, including existing and under construction lines (see Figure \ref{fig:topology and demand}). High Voltage Direct Current (HVDC) is assumed for the transmission lines, whose capacities can be expanded by the model if it is cost-effective. Within each country, different buses are connected by energy converters as shown in Figure \ref{fig:flow}.

The model covers both electricity and heating sectors, with converters coupling them in order to utilise the flexibility in both sectors. It runs over a full year of hourly data. The inelastic electricity and heating loads, including the ratio between urban and rural heating, are exogenous to the model and not optimised, as well as for hydroelectricity, \textit{i.e.}, hydro reservoir, run-of-river generators, and pumped-hydro storage, where fixed capacities are assumed due to environmental concerns. By contrast, VRES generator capacities, \textit{i.e.}, onshore wind, offshore wind, and solar PV, conventional generator capacities, \textit{i.e.}, open cycle gas turbines (OCGT), combined heat and power (CHP), gas boilers, converter capacities, \textit{i.e.}, heat pumps and resistive heaters, storage power and energy capacities, \textit{i.e.}, batteries and hydrogen for electricity and hot water tanks for heating, and transmission capacities are all optimised. In addition, the hourly operational dispatch of generators, converters, and storage units are subject to optimisation as well.

\subsection{Objective function}
For each country $i$, there are three buses: electricity, urban heat and rural heat. See Figure \ref{fig:flow}. All buses are labelled by $n$, generators and storage technologies by $s$, hour of the year by $t$, and bus connectors by $\ell$, which include transmission lines and converters. The total annual system cost consists of fixed annualised costs $c_{n,s}$ for generator and storage power capacity $G_{n,s}$, fixed annualised costs $\hat{c}_{n,s}$ for storage energy capacity $E_{n,s}$, fixed annualised costs $c_\ell$ for bus connectors $F_{\ell}$, variable costs, including \co{} tax $o_{n,s,t}$, for generation and storage dispatch $g_{n,s,t}$. The total annual system cost is minimised by:
\begin{align}
& \min_{\substack{G_{n,s},E_{n,s},\\F_\ell,g_{n,s,t}}} \left[ \sum_{n,s} c_{n,s} \cdot G_{n,s} +\sum_{n,s} \hat{c}_{n,s} \cdot E_{n,s} \right. \nonumber \\
& \hspace{2cm} \left. + \sum_{\ell} c_{\ell} \cdot F_{\ell}+ \sum_{n,s,t} o_{n,s,t} \cdot g_{n,s,t} \right]
\label{eq:objective}
\end{align}

The Levelised Cost of Electricity is usually employed to calculate the cost of the electricity generated by different technologies \cite{Ueckerdt_2013} or power systems \cite{Eriksen2017} in a consistent manner. We extend that definition here and define the Levelised Cost of Energy (LCOE) as the total system cost per unit of consumed energy, that is, including supplied electricity and heating demand. 

\subsection{Constraints}
The demand $d_{n,t}$ of bus $n$ at hour $t$ is met by VRES generation, hydroelectricity, conventional backup (OCGT, CHP, gas boiler), storage discharge, converters (heat pump, resistive heater) and HVDC transmission across border.
\begin{equation}
\sum_{s} g_{n,s,t}+ \sum_{\ell} \alpha_{n,\ell,t}\cdot f_{\ell,t} = d_{n,t} \hspace{.2cm} \leftrightarrow \hspace{0.2cm} \lambda_{n,t} \hspace{.3cm} \forall\, n,t \label{eq:energybalance}
\end{equation}
where $f_{\ell,t}$ refers to energy flow on the link $l$ and $\alpha_{n,\ell,t}$ indicates both the direction and the efficiency of flow on the bus connectors; it can be time-dependent such as heat pumps. The Lagrange/Karush-Kuhn-Tucker (KKT) multiplier $\lambda_{n,t}$ associated with the demand constraint represents the local marginal price of the energy carrier. 

The dispatch of generators and storage is bounded by the product between installed capacity $G_{n,s}$ and availabilities $\ubar{g}_{n,s,t}$, $\bar{g}_{n,s,t}$:
\begin{equation}
\ubar{g}_{n,s,t} \cdot G_{n,s} \leq g_{n,s,t} \leq \bar{g}_{n,s,t} \cdot G_{n,s} \hspace{1cm} \forall\, n,s,t \; . \label{eq:gen}
\end{equation}
$\ubar{g}_{n,s,t}$ and $\bar{g}_{n,s,t}$ are time-dependent lower and upper bounds due to, \textit{e.g.}, VRES weather-dependent availability. For instance, for wind generators, $\ubar{g}_{n,s,t}$ is zero and $\bar{g}_{n,s,t}$ refers to the capacity factor at time $t$. $G_{n,s}$ is the installed power capacity for generators, limited by installable potentials $\bar{G}_{n,s}$ due to, \textit{e.g.}, geographical constraints:
\begin{equation}\label{eq:geo limit}
0 \leq G_{n,s}\leq \bar{G}_{n,s} \hspace{1cm} \forall\, n,s \; .
\end{equation}
Similarly, the dispatch of converters has to fulfil the following constraints
\begin{equation}
\ubar{f}_{\ell,t} \cdot F_{\ell} \leq f_{\ell,t} \leq \bar{f}_{\ell,t} \cdot F_{\ell} \hspace{1cm} \forall\, \ell,t \; . \label{eq:con}
\end{equation}
For a unidirectional converter, \textit{e.g.}, a heat pump, $\ubar{f}_{\ell,t}=0$ and $\bar{f}_{\ell,t}=1$ since a heat pump can only convert electricity into heat. For transmission links, $\ubar{f}_{\ell,t}=-1$ and $\bar{f}_{\ell,t}=1$, which allows both import and export between neighbouring countries. In particular, the inter-connecting transmission can be limited by a global constraint
\begin{equation}
\sum_{\ell} l_\ell \cdot F_{\ell} \leq  \textrm{CAP}_{LV} \hspace{.7cm} \leftrightarrow \hspace{0.3cm} \mu_{LV} \; ,
%\hspace{.3cm} 
\label{eq:lvcap}
\end{equation}
where the sum of transmission capacities $F_{\ell}$ multiplied by the lengths $l_{\ell}$ is bounded by a transmission volume cap $\textrm{CAP}_{LV}$. The KKT multiplier $\mu_{LV}$ associated with the transmission volume constraint indicates the shadow price of an increase in transmission volume to the system. 

The state of charge $e_{n,s,t}$ of every storage has to be consistent with charging and discharging in each hour, and is limited by the energy capacity of the storage $E_{n,s}$
\begin{align}
e_{n,s,t} = & \ \eta_0 \cdot e_{n,s,t-1} + \eta_{1} |g_{n,s,t}^+| - \eta_{2}^{-1} |g_{n,s,t}^-| \nonumber \\
& + g_{n,s,t,\textrm{inflow}} - g_{n,s,t,\textrm{spillage}} \; , \nonumber \\
& 0  \leq   e_{n,s,t} \leq E_{n,s}   \hspace{0.5cm} \forall\, n,s,t \; . \label{eq:soc}
\end{align}
The storage has a standing loss $\eta_0$, a charging efficiency $\eta_1$ and rate $g_{n,s,t}^+$, a discharging efficiency $\eta_2$ and rate $g_{n,s,t}^-$, possible inflow and spillage which are subject to Equation (\ref{eq:gen}). The storage energy capacity $E_{n,s}$ can be optimised independently of the storage power capacity $G_{n,s}$.

To enforce the decarbonisation of the energy system, Brown \textit{et al.} \cite{Brown_2018} imposed a 95\% \co{} emission reduction constraint relative to 1990. We follow a different approach here. No \co{} constraint is imposed but the impact of variable \co{} prices, ranging from 0 to 500~\euro/t\co{} is investigated. %In addition, an unrealistic extremely high \co{} price (10,000~\euro/t\co{}) is considered to identify the most efficient coupled system.

\subsection{VRES layout} \label{sec:VRES_layout}
The VRES layout is considered at a country level instead of for each bus. For every country $i$, the annual available VRES generation is denoted by $g_{i,VRES}^{gross}$, where `gross' represents the energy that can be potentially generated, that is, before curtailment.
\begin{equation*}
g_{i,VRES}^{gross} = \sum_{t,s\in VRES,n\in i} \bar{g}_{n,s,t} \cdot G_{n,s}
\end{equation*}
The gross VRES penetration $\gamma_i^{gross}$ is defined as the ratio of gross VRES generation to the total demand in country $i$, which is the sum of electricity and heating demands
\begin{equation}
g_{i,VRES}^{gross} = \gamma_i^{gross} \sum_{t,n\in i} d_{n,t} \; .
\end{equation}
The gross VRES generation consists of gross wind generation $g_{i,W}^{gross}$ and gross solar generation $g_{i,S}^{gross}$,
\begin{equation}
g_{i,VRES}^{gross} = g_{i,W}^{gross}+g_{i,S}^{gross} \; .
\end{equation}
The gross wind/solar mix parameter $\alpha_i^{gross}$ determines the ratio between available wind and VRES
\begin{equation}
g_{i,W}^{gross} = \alpha_i^{gross} g_{i,VRES}^{gross} \; .
\end{equation}
The gross VRES mix of the whole system $\alpha_{EU}^{gross}$ can be found by
\begin{equation}
\sum_{i} g_{i,W}^{gross} = \alpha_{EU}^{gross} \sum_{i} g_{i,VRES}^{gross} \; ,
\end{equation}
where $\alpha_{EU}^{gross}$ expresses the overall VRES layout tendency towards wind or solar dominance. 

To utilise different VRES resources over the continent, a weakly homogeneous layout \cite{Eriksen2017} is introduced, where `homogeneous' indicates that the share $\gamma_i^{gross}$ of each country is the same, which can be shortened to $\gamma^{gross}$, and `weakly' suggests that the mix $\alpha_i^{gross}$ of each country is optimised. 
\begin{equation}
\gamma_i^{gross} = \gamma^{gross}, \hspace{0.2cm} \alpha_i^{gross} \textrm{ subject to opt.}
\end{equation}
This layout ensures that each country is VRES self-sufficient to a certain extent, and the optimisation seeks the optimal gross wind/solar mix in each country. It should be remarked that this implementation is different to that in \cite{Brown_2018} where nodal VRES self-sufficiency might not be fulfilled.

%It is also useful to define the annual generated VRES $g_{i,VRES}^{net}$, where `net' represents the energy that has been generated, after curtailment. 
%\begin{equation*}
%g_{i,VRES}^{net} = \sum_{t,s\in VRES, n\in i} g_{n,s,t} \; .
%\end{equation*}
%The net VRES penetration is the ratio of generated VRES to the total demand
%\begin{equation}
%g_{i,VRES}^{net} = \gamma_i^{net} \sum_{t,n\in i} d_{n,t} \; .
%\end{equation}
%The European net VRES penetration $\gamma_{EU}^{net}$ can be calculated by 
%\begin{equation}
%\sum_{i} g_{i,VRES}^{net} = \gamma_{EU}^{net} \sum_{i} \sum_{t,n\in i} d_{n,t} \; .
%\end{equation}

%-----------------------------
\section{Methods: Data} \label{sec:methods data}
%-----------------------------

A brief description of the input data is provided below. The reader is referred to \cite{Brown_2018} for a comprehensive explanation of the technologies included in the model, their costs, efficiencies, and additional parameters. 

\subsection{Countries and network}

The network considered has been previously described in \cite{Schlachtberger_2017}. It comprises 30 nodes, each of them represents a country of the 28 European Union member states as of 2018 excluding Malta and Cyprus but including Norway, Switzerland, Serbia and Bosnia-Herzegovina. The nodes are connected through links representing the current and planned transmission lines between countries. See Figure \ref{fig:topology and demand}. 

\subsection{Electricity and heating demand}

Electricity and heating demands are modelled using hourly time series for every node, \textit{i.e.}, country in the network. For the electricity demand, historical values corresponding to the year 2015 are gathered from the European Network of Transmission System Operators for Electricity (ENTSO-e) \cite{ENTSOE} through the very convenient unified file provided by the Open Power System Data (OPSD) project \cite{OPSD}. The electricity demand time series for each country is computed by subtracting the domestic space and water heating demands, which are included as separate demand time series (described in the next paragraph). See Figure \ref{fig:topology and demand} for demand profiles.

For the heating demand, only residential and service space heating and hot water demands are included. In this case, temperature data from reanalysis, \textit{i.e.}, the outcome of global climate models, and the Heating Degree Hour (HDH) approximation are used \cite{Quayle_1979}. The Climate Forecast System Reanalysis (CFSR) dataset from the National Center for Environmental Prediction (NCEP) \cite{CFSR} which includes temperature values at high spatial and time resolution (0.3125$^\circ \times 0.3125^\circ$, 1 hour) is used as input. For Europe, the spatial resolution translates into a grid point every $40\times40~km^2$ approximately. 

The HDH approach assumes that the heating demand increases linearly from a threshold temperature of \SI{17}{\celsius}. HDH is computed for every point in the CFSR grid, weighted by the population density (the NUTS3 population data \cite{NUTS3} is used as a proxy) and aggregated at a national scale to obtain HDH time series representative for a country. The time series is then scaled based on the annual demand for domestic space heating in 2015, which is retrieved from the Heat Roadmap Europe project \cite{HRE}. Finally, a constant hourly value for the hot water consumption, obtained from the same database, is added to compute the total heating demand time series representative for every country.

The estimated values for total annual demand in Europe are similar for electricity and heating, accounting for 2,854 TWh\el/a and 3,562 TWh\th/a respectively.

\subsection{Electricity generation}
The model for electricity generation was thoroughly described in \cite{Schlachtberger_2017} and only a general overview is provided here. In the electricity bus, the following generation technologies are considered: onshore wind, offshore wind, solar PV, hydro reservoirs, run-of-river generators, Open Cycle Gas Turbines (OCGT) and Combined Heat and Power (CHP) units. The inflow for hydro reservoirs and run-of-river generators are considered fixed to 2011 values, with an annual inflow of 473 TWh\el. Pumped hydro storage and hydro reservoirs are able to store electricity, and their power capacities and energy capacities are fixed to 2011 values, \textit{i.e.}, 170 GW and 204 TWh\el, respectively. A round-trip efficiency of 75\% (0.866$\cdot$0.866) has been assumed for the pumped hydro storage. The sources for all hydro-related data are included in \cite{Brown_2018}.

For wind and PV, time series of hourly capacity factors representative for every country are computed using the Global Renewable Energy atlas (REatlas) from Aarhus University \cite{Andresen2015, victoria2018using}. The REatlas uses the CFSR dataset as input. Simplified models of wind turbines and PV panels are employed to transform wind velocity, irradiance, and temperature from the reanalysis dataset into time-dependent electricity generation. Time series for every CSFR grid points are aggregated at a national level to obtain hourly capacity factors representative for every country. For wind time series, a smoothing function is applied to wind turbines whose parameters have been optimised for every country using historical data. A detailed description of the methodology and its validation can be found in \cite{Andresen2015}. For solar PV, reanalysis irradiance is bias corrected using SARAH satellite database to include the effect of local atmospheric characteristics in every country. The methodology and its validation based on PV historical time series can be found elsewhere \cite{Victoria_2018}.

\begin{table*}[!t]
	\begin{threeparttable}
		%\captionsetup{labelformat=empty}
		\caption{Key cost parameters} \label{tab:cost parameters}
		\centering
		\begin{tabularx}{0.95\textwidth}{lcccccc}
			\toprule
			Technology                 &Overnight   &Unit &FOM\tnote{a} &Lifetime &CF\tnote{b}/ & LCOE\tnote{b} 		\\
			&Cost[\euro] &     &[\%/a] 		&[a]   	  &Efficiency   &[\euro/MWh] 			\\
			\midrule
			Onshore wind\tnote{c}             &910         &kW\el  &3.3 &30  &0.23[0.07-0.33]	&52[35-224]    \\
			Offshore wind\tnote{c}             &2506        &kW\el  &3   &25  &0.31[0.09-0.51]	&91[66-182] 	\\
			Solar PV\tnote{c}  		   &575         &kW\el  &2.5 &25  &0.13[0.06-0.19]	&55[39-114] 	\\
			OCGT\tnote{d}              &560         &kW\el  &3.3 &25  &0.39        				&63                     \\
			CHP \tnote{d}       	   &600         &kW\th  &3.0 &25  &0.47        				&54                     \\
			Gas boiler\tnote{d,e}      &175/63      &kW\th  &1.5 &20  &0.9         				&25/26 				    \\
			Resistive heater           &100         &kW\th  &2   &20  &0.9         				& -			            \\
			Heat pump\tnote{e}         &1400/933    &kW\th  &3.5 &20  &[3.03-3.79]/[2.73-3.04]  & -	  		\\
			Battery storage\tnote{f}   &144.6		&kWh    &0   &15  &1.0		   				& - 					\\
			%Battery converter		   &310			&kW\el  &3   &20  &$0.9\cdot0.9$ 			& - 				   	\\
			Hydrogen storage\tnote{f}  &8.4			&kWh	&0	 &20  &1		   				& -					    \\
			Hot water tank\tnote{e,f}	   &860/30		&$m^3$  &1	 &20/40 &$\tau=3/180$ days		& - 				    \\
			HVDC lines	  			   &400			&MWkm	&2	 &40  &1		   				& - 					\\
			\bottomrule
		\end{tabularx}
		
		\begin{tablenotes}
			\footnotesize
			\item [a] Fixed Operation and Maintenance (FOM) costs are given as a percentage of the overnight cost per year.
			\item [b] Capacity Factor (CF) only applies to renewables, and efficiency only applies to generators and converters. LCOE is calculated assuming a discount rate of 0.07.
			\item [c] Capacity Factor varies in different countries due to weather condition, hence influences the LCOE of VRES. The number in front indicates the average of CF/LCOE weighted by demand, while the numbers in brackets show the range of CF/LCOE for different countries. Solar PV is split 50-50\% between rooftop and utility.
			\item [d] The fuel cost of OCGT, CHP, and gas boiler is 21.6~\euro/MWh\th. The efficiency and LCOE of CHP are calculated assuming they work in the condensing mode \cite{Brown_2018}.
			\item [e] Gas boilers, heat pumps, and hot water tanks have different costs and time constant $\tau$ for decentralised (numbers in front) and centralised (numbers behind) systems. The efficiency of heat pumps, also known as the Coefficient of Performance (COP), varies with temperature.
			\item [f] The overnight costs of storage converters, such as battery charge and discharge, are not included in the storage costs.
		\end{tablenotes}
	\end{threeparttable}
\end{table*}

The maximum capacities of every technology that can potentially be installed in every country are constrained by several limits. Protected sites listed in Natura 2000 \cite{Natura2000}, as well as areas with certain land use types, \textit{e.g.}, cities, as specified by \cite{Corine_2014}, are excluded. For offshore wind, the maximum water depth assumed is 50 m. On top of these, we assumed that only 20\% of the resulting available area can be used for wind and 1\% for PV, which results in a maximum density of 2 MW/km$^2$ for wind and 1.5 MW/km$^2$ for PV.
%also limits the capacities that can be installed at every point in the CFSR grid. 
Regarding PV, the model assumes that 50\% of the installed capacities correspond to rooftop mounted systems while the other 50\% belongs to larger utility-scale power plants with different costs assumed for every type of installation \cite{Brown_2018}. 

\subsection{Heating supply}
The technologies available in the model that supply heat include gas boilers, resistive heaters, heat pumps and CHP units. A lower cost can be achieved if those conversion technologies are built at a larger scale to feed-in district heating systems. Since centralised solutions are only cost-effective when the population density is above a certain threshold, two heat buses are considered in the model (see Figure \ref{fig:flow}), where the overall heating demand is divided proportionally into \textit{urban heat} and \textit{rural heat}. The \textit{urban heat} bus supplies heating demand in places whose population density is large enough to allow district heating. The cost-optimal solution can include district heating for every country in Europe except for southern countries (Spain, Greece, Portugal, Italy and Bulgaria) where the high winter temperatures discourage the construction of district heating systems. Where district heating is allowed, CHP plants can be built to feed-in the system. Central ground-sourced heat pumps have been assumed in urban areas with district heating, while air heat-sourced pumps have been assumed for individual systems in urban areas. On the other hand, the \textit{rural heat} bus represents the places where only decentralised heating units are allowed. In this case, individual ground-sourced heat pumps have been assumed. Despite being more expensive, the higher COP of ground-sourced heat pumps makes them economically favourable. Additionally, in both heat buses, resistive heaters or gas boilers can also be used to supply the demand. The description of the models assumed for the temperature-dependent efficiency of heat pumps and possible electricity-heat output combinations for CHP, as well as the efficiencies assumed for gas boilers and resistive heaters, are included in \cite{Brown_2018} and summarised in Table \ref{tab:cost parameters}.

\subsection{Storage}
Beside pumped hydro storages, whose energy and power capacities are fixed exogenously to the model, three additional technologies are included in the model: short-term electric batteries, long-term hydrogen storage and thermal energy storage. In the electricity bus, the energy and charging capacities of batteries and $H_2$ storage are independently optimised. In the heat buses, thermal energy storages, \textit{i.e.}, water tanks, can be built. In the urban heat bus of northern countries where the district heating solution is allowed, long-term storage representing large water tanks connected to district heating systems can be built. A time constant $\tau=180$ days is assumed for these long-term storages, \textit{i.e.}, $1-\exp(-\frac{1}{24\tau})$ of the stored energy is lost per hour regardless of the ambient temperature. Where only decentralised solutions are allowed, only individual thermal energy storages can be built. They have a significantly lower time constant, $\tau=3$ days, that makes them more suitable for short-term storage.

\begin{figure*}[t!]
	\centering
	\includegraphics[trim=0 0cm 0 0cm,width=\linewidth,clip=true]{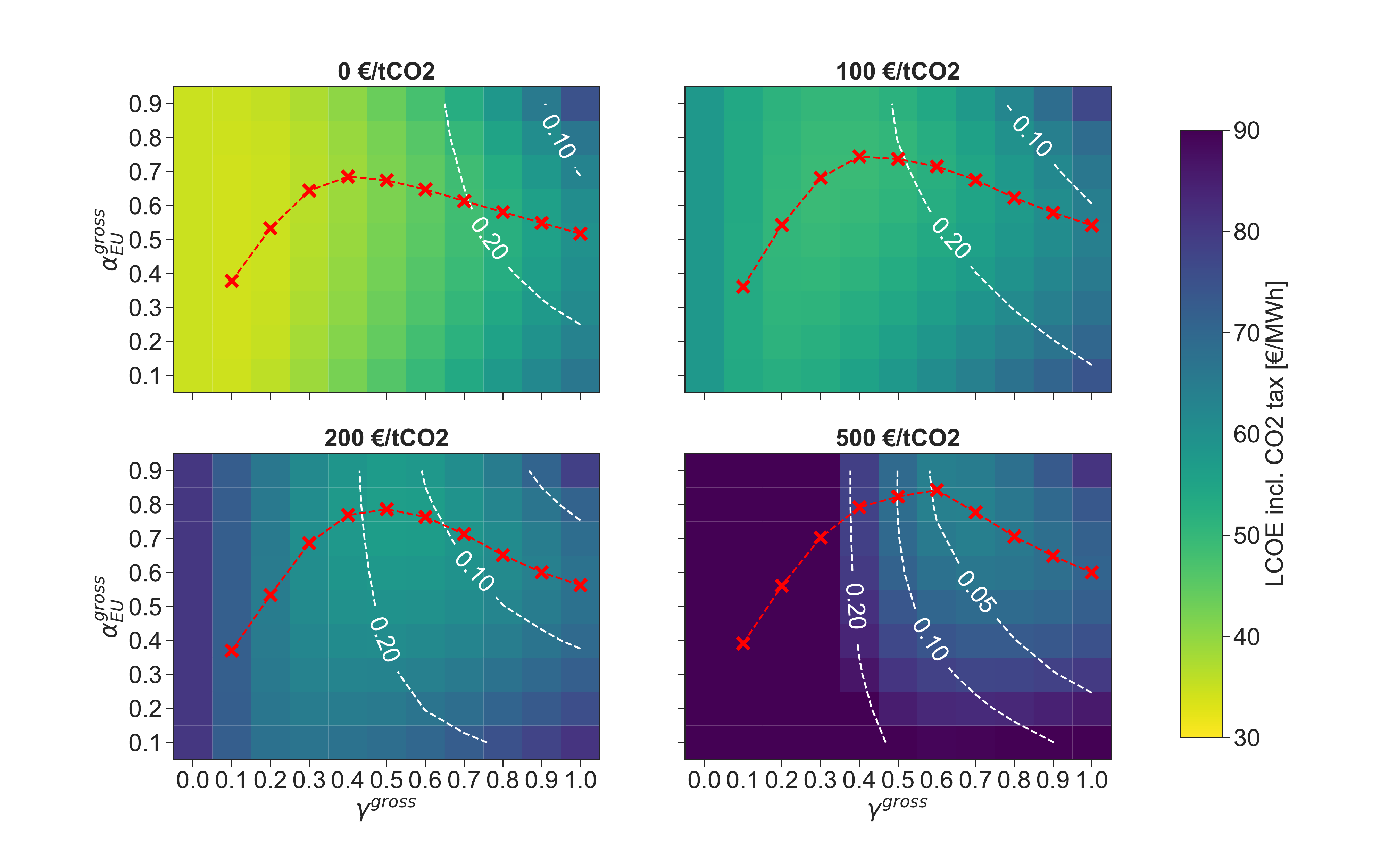}
	\caption{Levelised Cost Of Energy (LCOE) including \co{} tax, in \euro/MWh, as a function of gross VRES penetration $\gamma^{gross}$ and gross wind/solar mix $\alpha_{EU}^{gross}$ for different \co{} price cases (0, 100, 200, 500~\euro/t\co{}), assuming optimal transmission and sector coupling. Red crosses indicate the cost-optimal gross wind/solar mix of the corresponding VRES penetration. The white dotted lines depict the contours of 20\%, 10\% and 5\% \co{} emissions relative to 1990 level.}
	\label{fig:sweep}
\end{figure*}

\subsection{Costs and CO$_2$ emissions}
The annualised investment and variable costs for the different technologies, their lifetimes, as well as the assumed fuel cost, are summarised in Table \ref{tab:cost parameters}. They are mostly based on predictions for 2030 assuming a discount rate of 7\%. 

The CO$_2$ emissions in the model come only from those technologies using gas (OCGT, CHP and gas boilers). An emission factor equal to 0.19 tCO$_2$/MWh$_{th}$ has been assumed for gas. Coal is not considered in this paper.

%--------------------
\section{Results}\label{sec:result}
%--------------------
This section is organised as follows. First, the combined effect of VRES penetration, gross wind/solar mix, and \co{} pricing on the system costs and \co{} emissions is investigated in Section \ref{subsec:configuration}. Second, the target configurations, defined as the cost-optimal configurations to achieve 5\%, 10\%, and 20\% \co{} emissions, relative to 1990 level, are identified in Section \ref{subsec:target}. Third, the characteristics of the three target configurations are analysed in Section \ref{subsec:distribution} for the 30 European countries comprising the model.

\subsection{System costs and \co{} emissions for different configurations} \label{subsec:configuration}
Figure \ref{fig:sweep} shows the LCOE including \co{} tax as a function of gross VRES penetration $\gamma^{gross}$ and gross wind/solar mix $\alpha_{EU}^{gross}$ for four different \co{} prices, assuming optimal transmission and sector coupling. The LCOE is calculated as the total system cost per unit of consumed energy, that is, including electricity and heating demand. Maximum $\gamma^{gross}$ investigated is 1.0 since higher values will lead to significant VRES curtailment. $\alpha_{EU}^{gross}$ ranges from 0.1 to 0.9. $\alpha_{EU}^{gross}=0$ and $\alpha_{EU}^{gross}=1$ indicates solar-only and wind-only respectively. These extremes have been excluded since the \WH constraint together with the geographical limitations makes the model infeasible. 

Note first that the evolution of the cost-optimal $\alpha_{EU}^{gross}$ as a function of $\gamma^{gross}$ (marked with red crosses in Figure \ref{fig:sweep}) shows a similar concave shape regardless of \co{} price. This indicates that the \co{} price has limited impact on the cost-optimal gross wind/solar mix. Although solar is slightly preferred at $\gamma^{gross}=0.1$, more wind is deployed as the VRES penetration increases until a maximum $\alpha^{gross}_{EU}$ is reached. The first stage of increasing optimal gross wind/solar mix is mainly due to the characteristics of solar PV generation. Since PV generation shows a strong daily pattern, a large short-term storage energy capacity is needed to counterbalance the diurnal variation. For low $\gamma^{gross}$, the PV daily cycle only affects a small share of the total generation and solar electricity can be easily integrated. For higher $\gamma^{gross}$, by contrast, a wind/solar mix favouring wind becomes easier to integrate because less storage is required and system cost is minimised. The optimal wind/solar mix reaches the maximum value when the best sites for onshore wind have been exploited. Then, as gross renewable penetration continues to raise, more solar is employed because some countries have reached their geographical limits for onshore wind capacity.

%Figure \ref{fig:LCOE} is obtained by extracting the cost-optimal gross wind/solar mix, which is marked by red crosses. The top plot shows the \co{} emissions while the bottom depicts the LCOE including/excluding \co{} price. 

The white dashed lines in Figure \ref{fig:sweep} mark the borders of the regions where 5\%, 10\%, and 20\% \co{} emissions, relative to 1990, are attained. For the plots in Figure \ref{fig:sweep}, except for the 500~\euro/t\co{} case, the region with the lowest prices and that with the lowest \co{} emissions do not overlap. For a zero \co{} price, the cost-optimal configuration lies at low VRES penetration ranges, while the highest \co{} reduction is attained at high VRES penetration. As \co{} price increases, the most cost-effective $\gamma^{gross}$ moves towards higher values, and the area of the plot where \co{} emissions become lower than 20\% extends towards lower $\gamma^{gross}$. Adding a \co{} price makes VRES resources more competitive and reduces the \co{} emissions significantly. In Figure \ref{fig:sweep}, this is illustrated by the while contours moving towards lower values of $\gamma^{gross}$ for increasing \co{} prices. Nonetheless, a \co{} price equal to 100~\euro/t\co{} does not manage to reduce the emissions of the coupled system to 5\% of 1990 level, even at a gross VRES penetration of 1.0. For the 200~\euro/t\co{} case, it is possible to bring \co{} emissions down to 5\%, but it is not compatible with the cost-optimal wind/solar mix. If \co{} price is increased to 500~\euro/t\co{}, the area including the cost-optimal configurations overlaps with the 5\% emission region. This indicates that high VRES penetrations do not necessarily lead to low \co{} emissions without \co{} tax.

\begin{figure}[!t]
	\centering	
	\includegraphics[trim=0 0cm 0 0cm,width=\linewidth,clip=true]{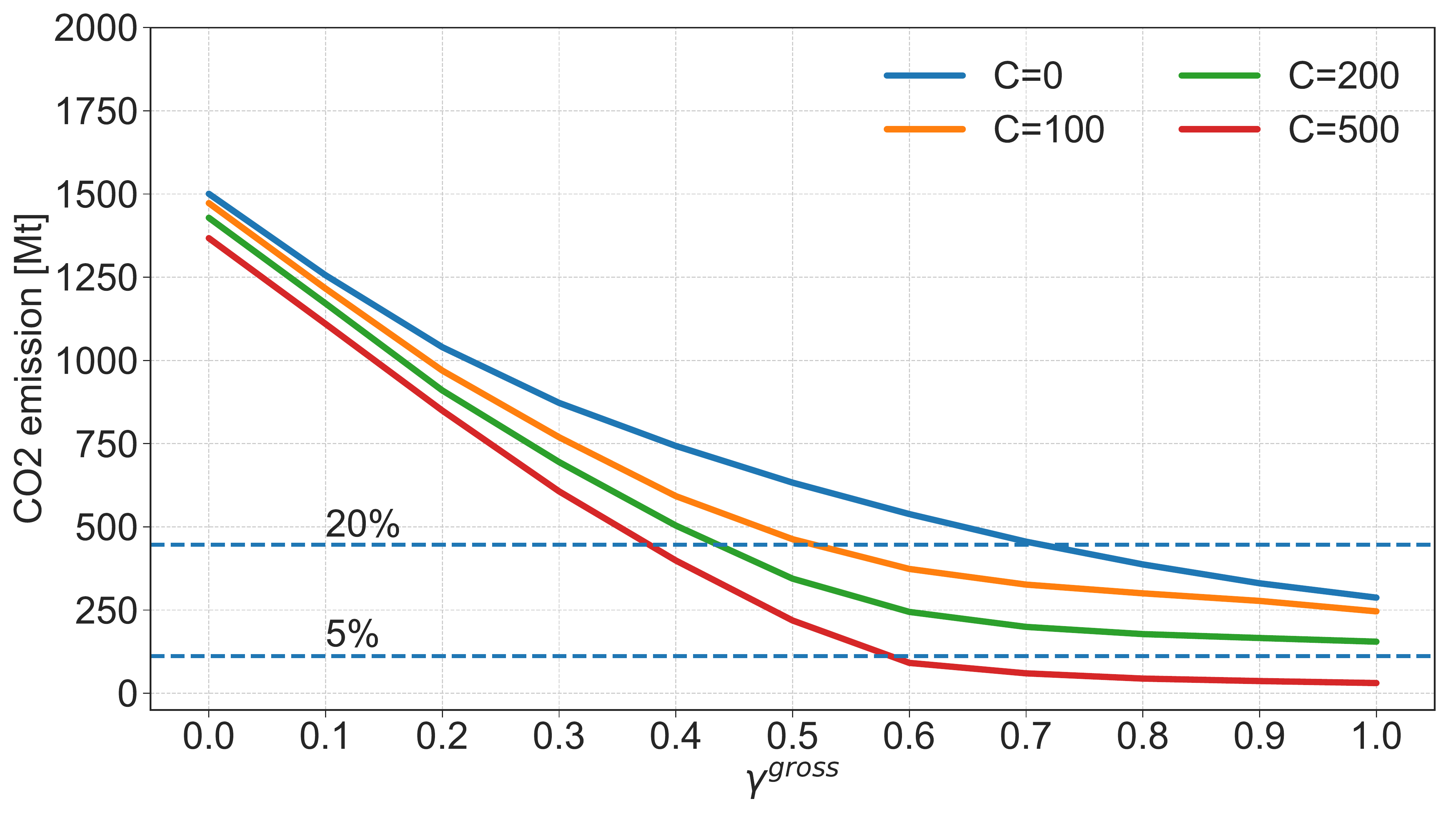}
	\includegraphics[trim=0 0cm 0 0cm,width=\linewidth,clip=true]{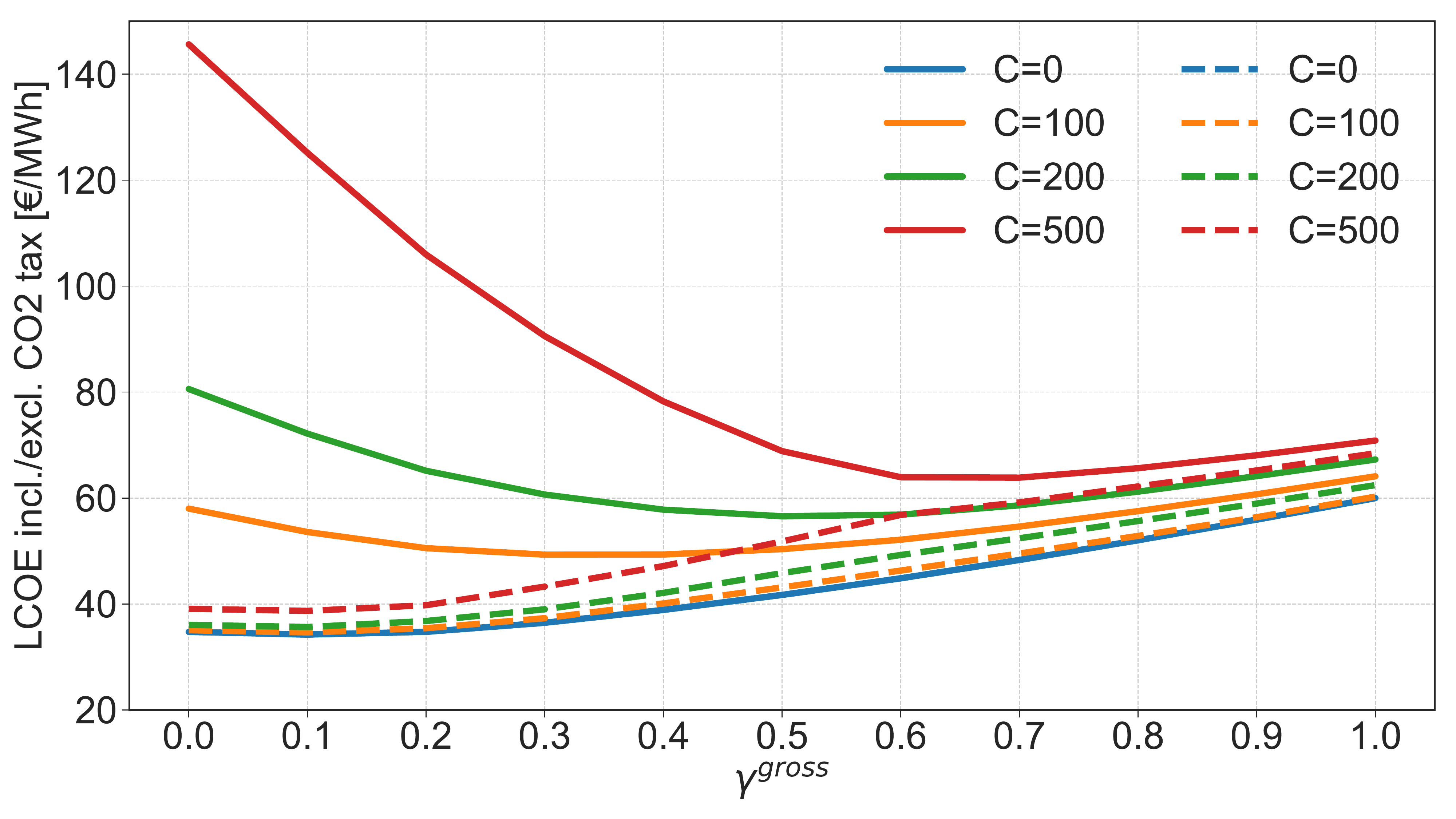}
	\caption{Top: \co{} emission as a function of gross VRES penetration for different \co{} prices, assuming optimal wind/solar mix. A magnification of this plot for low \co{} emission values is shown in Figure \ref{fig:target configuration}. Bottom: LCOE including/excluding (solid/dashed) \co{} tax as a function of gross VRES penetration for different \co{} prices (in units of \euro/t\co).}
	\label{fig:LCOE}
\end{figure}

The necessity of a high \co{} price to achieve ambitious \co{} reductions is also shown in the top plot of Figure \ref{fig:LCOE} where the \co{} emission is shown as a function of $\gamma^{gross}$ for optimum $\alpha^{gross}_{EU}$, \textit{i.e.}, the data corresponding to red crosses in Figure \ref{fig:sweep}. The bottom plot of Figure \ref{fig:LCOE} depicts the LCOE as a function of $\gamma^{gross}$ also assuming optimum $\alpha^{gross}_{EU}$. The LCOE is shown both including and excluding \co{} taxes. In practice, \co{} tax is essentially a charge imposed to disincentivise the use of gas. The collected money could be returned to the system, for example, to co-finance new VRES capacities or efficiency and saving policies. 

Figure \ref{fig:LCOE} illustrates that, for \co{} price of 500~\euro/t\co{}, gross penetration of 0.6 can bring the \co{} emission down to 5\%, while LCOE corresponds to 64~\euro/MWh including \co{} tax and 57~\euro/MWh excluding \co{} tax. As VRES become the dominant primary energy in the system, the difference between LCOE including and excluding \co{} tax declines significantly. Figure \ref{fig:LCOE} emphasises that a \co{} price in the order of a few hundred \euro/t\co{} is required to attain simultaneously a cost-effective and low-carbon electricity-and-heating system. 

A final observation from Figure \ref{fig:sweep} is that, for the same $\gamma^{gross}$, wind-dominated systems emit less \co{} than solar-dominated systems. Although onshore wind and solar PV have similar LCOE according to Table \ref{tab:cost parameters}, diurnal variation of solar generation need to be balanced. Instead of a considerable amount of costly short-term battery storage \cite{Rasmussen2012,tranberg2018flow}, more conventional technologies are deployed in the cost optimisation of a solar-dominated system. This results in higher \co{} emissions. The phenomenon is more pronounced in highly-decarbonised systems. For instance, at 500~\euro/t\co{} and $\gamma^{gross}=0.7$, the case of optimal wind/solar mix ($\alpha^{gross}_{EU}$=0.8) is not only cheaper than the case of $\alpha^{gross}_{EU}$=0.1, but also implies lower \co{} emissions. 

\begin{figure}[!t]
	\centering	
%	\small \textbf{0~\euro/t\co{}}
	\includegraphics[trim=0 0cm 0 0cm,width=\linewidth,clip=true]{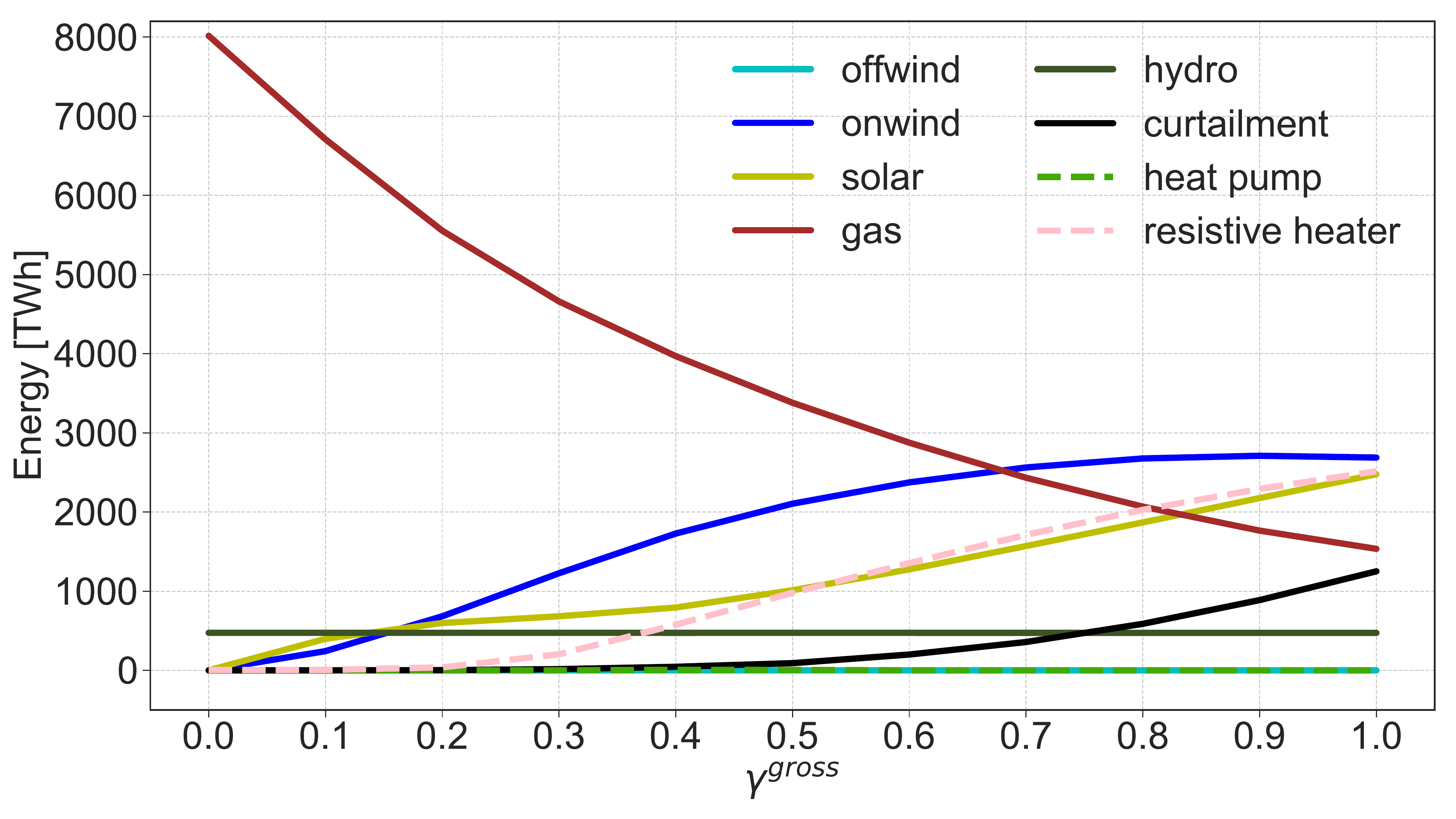}
%	\small \textbf{500~\euro/t\co{}}
	\includegraphics[trim=0 0cm 0 0cm,width=\linewidth,clip=true]{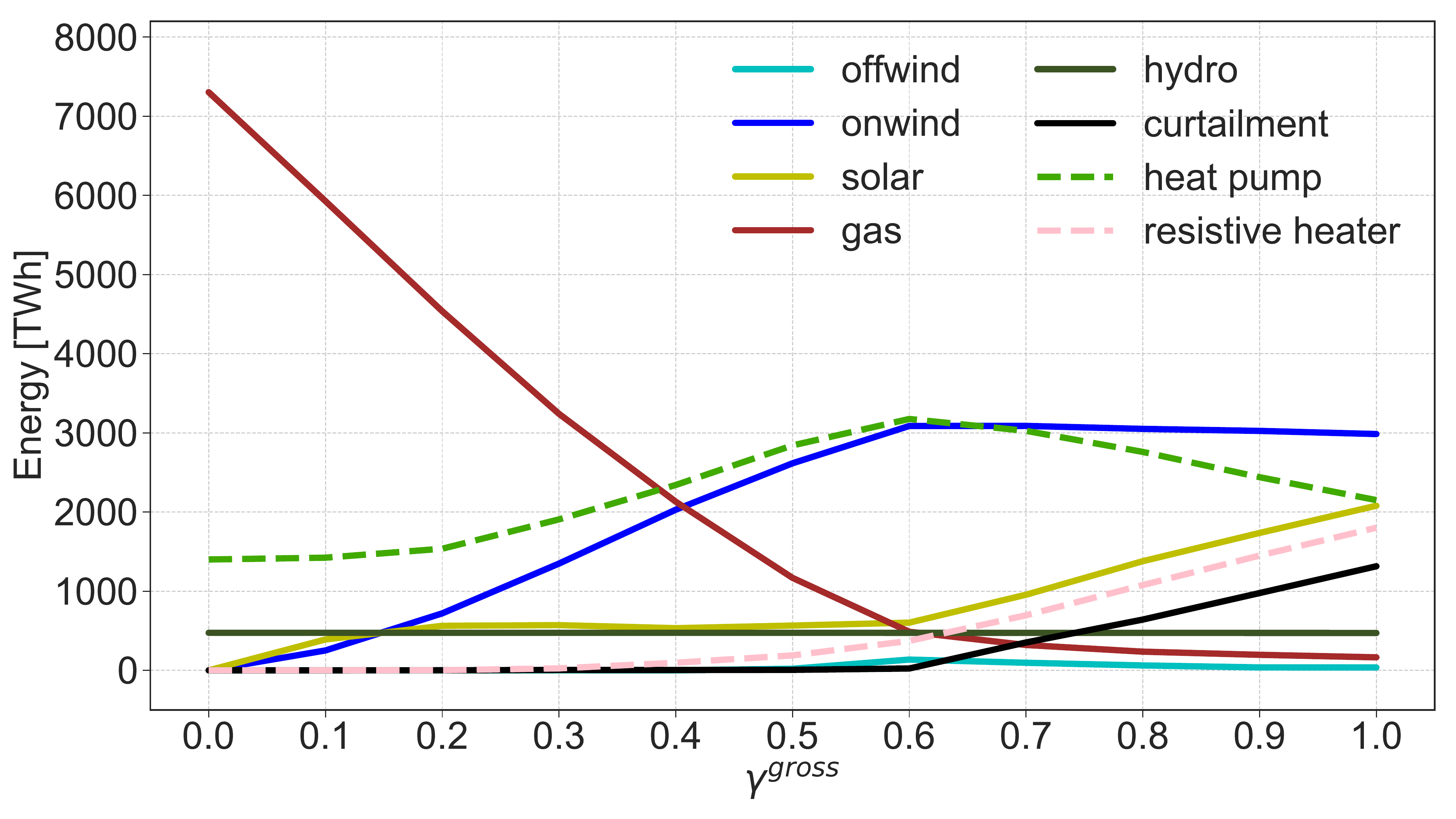}
	\caption{Primary, curtailed, and converted energy as a function of gross VRES penetration for \co{} price of 0~\euro/t\co (top) and 500~\euro/t\co (bottom), assuming optimal wind/solar mix. Primary energy consists of VRES, hydro, and gas, while curtailed energy only applies to VRES. Converted energy is depicted in dashed lines, comprising thermal energy produced by heat pumps and resistive heaters.}
	\label{fig:energy}
\end{figure}

While Figure \ref{fig:sweep} and Figure \ref{fig:LCOE} present the evolution of the system cost and \co{} emissions, Figure \ref{fig:energy} reveals the composition of generated, converted, and curtailed energy as a function of gross VRES penetration under two different \co{} price assumptions: zero and 500~\euro/t\co. A high \co{} price causes primary energy generated by gas to reduce more sharply and also incentivises the conversion of electricity to heat through heat pumps. Although VRES generation becomes more cost-competitive as the \co{} price increases, the VRES curtailment does not see a compelling diminution. One major reason is that system with high \co{} price utilises the primary energy more efficiently, \textit{e.g.}, heat pumps can provide more thermal energy compared to resistive heaters with the same amount of electricity consumption. 
% Marta: I deleted this line because it cannot be readed from the plot :"Comparing to 0~\euro/t\co price, 500~\euro/t\co{} with high gross penetration system demands less primary energy, thus failing to diminish the VRES curtailment."

\begin{figure}[!t]
	\centering
	\includegraphics[trim=0 0cm 0 0cm,width=\linewidth,clip=true]{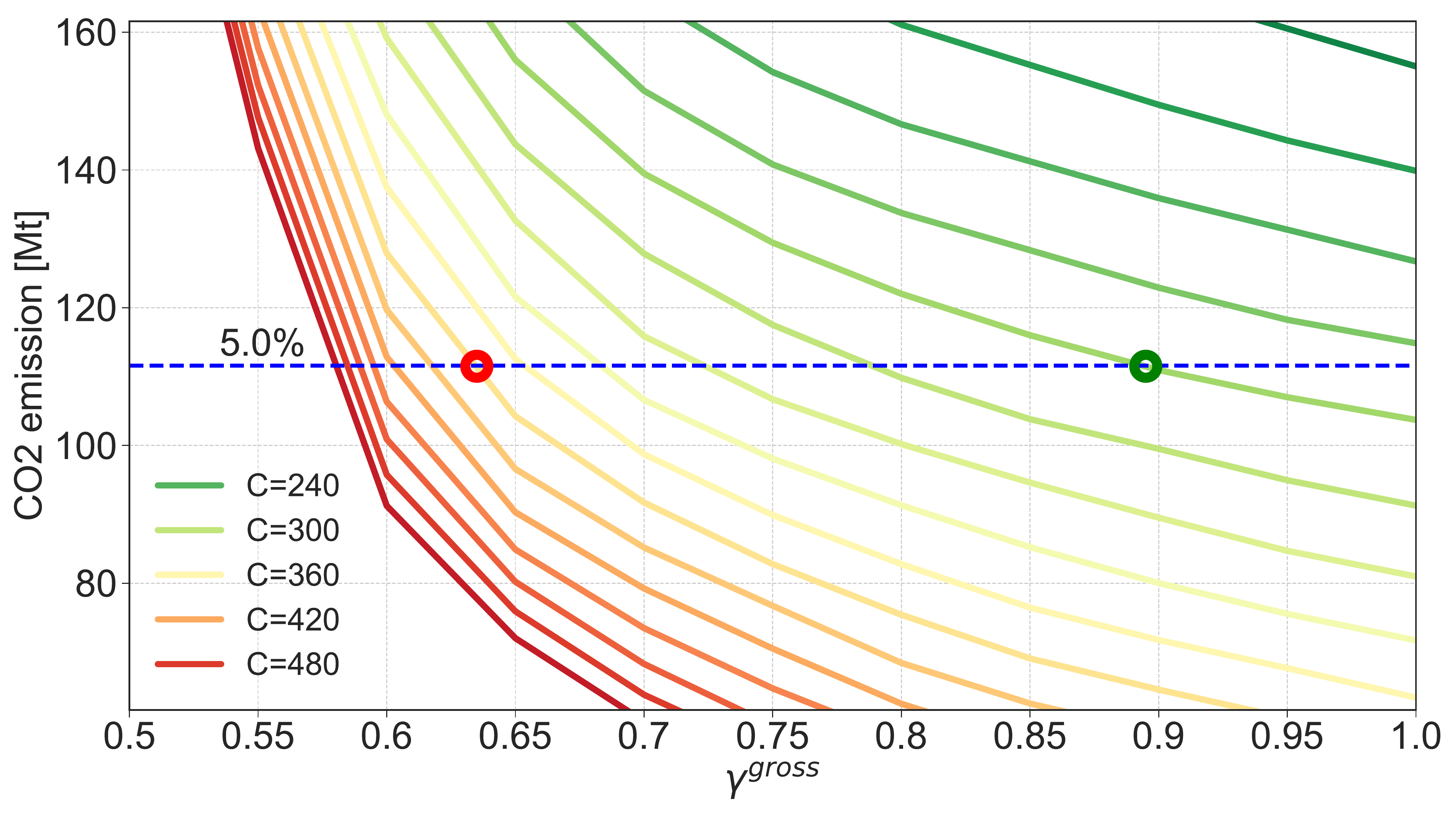}	
	\includegraphics[trim=0 0cm 0 0cm,width=\linewidth,clip=true]{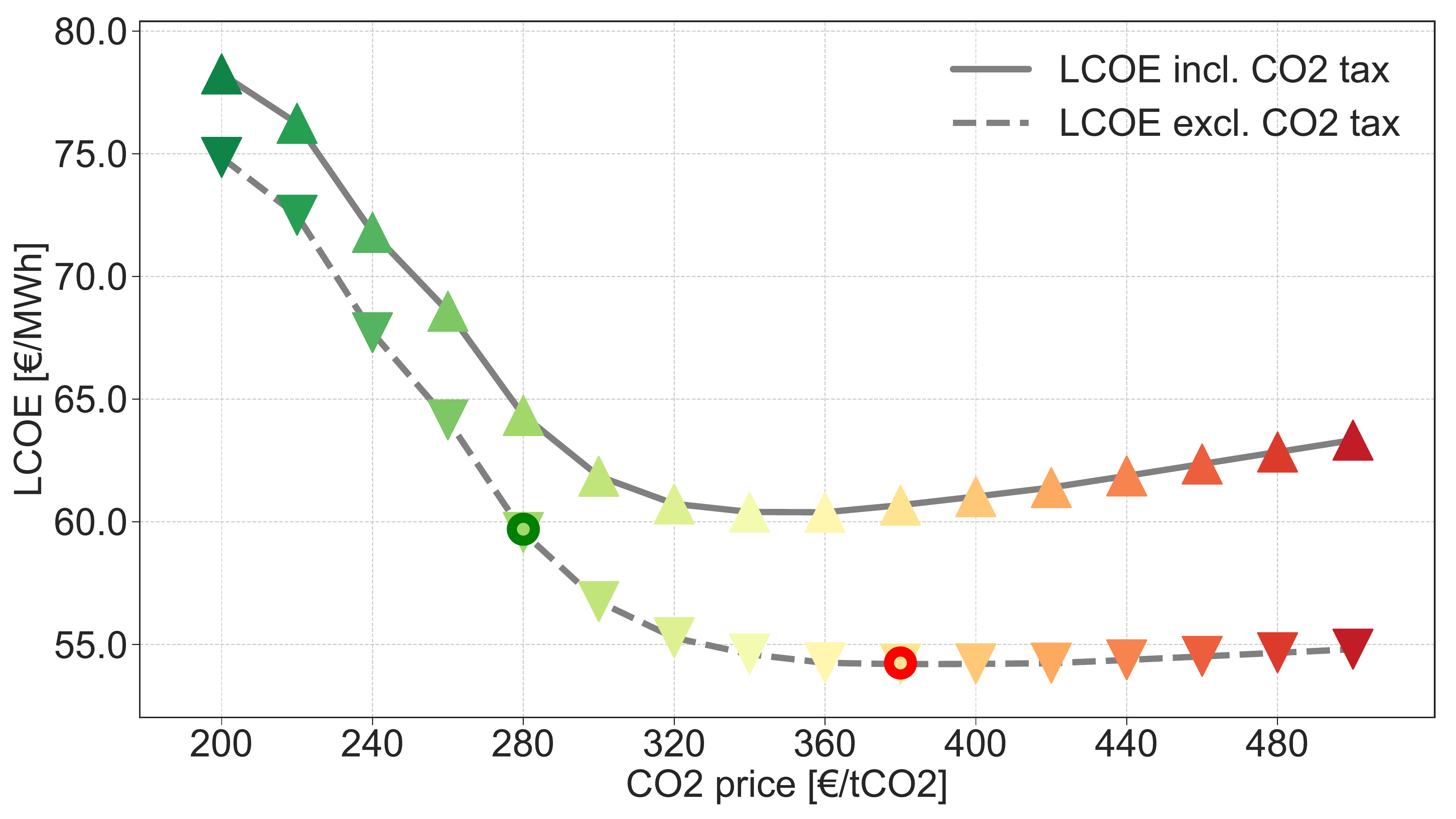}	
	\includegraphics[trim=0 0cm 0 0cm,width=\linewidth,clip=true]{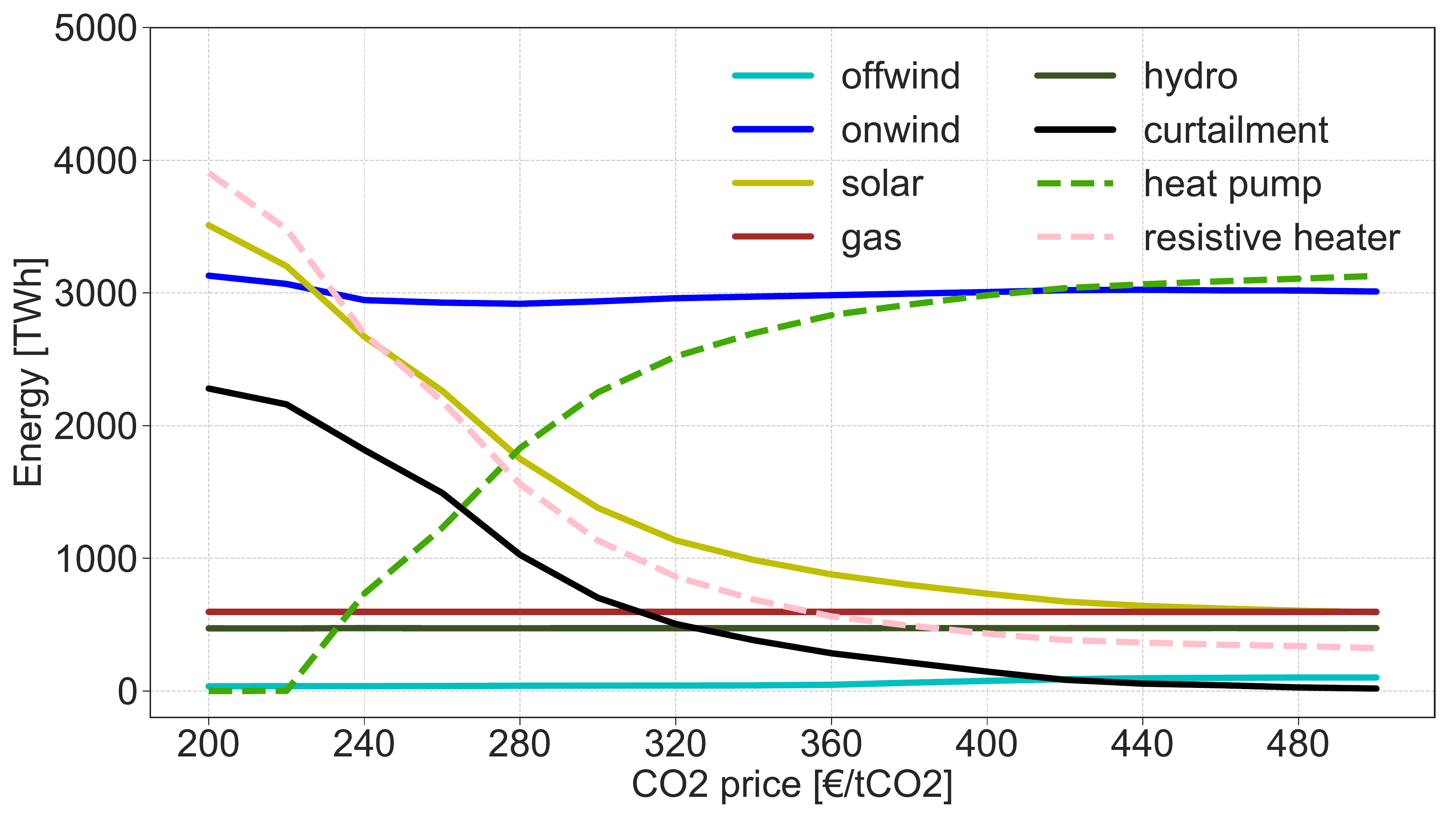}
	\caption{\co{} emissions (top) as a function of gross VRES penetration for different \co{} prices (in units of \euro/t\co{}), assuming optimal wind/solar mix. The blue dotted line represents 5\% of \co{} emission relative to 1990 level. For every \co{} price, the LCOE including and excluding \co{} tax intersecting the 5\% line is shown on the middle plot. Bottom plot depicts the primary, curtailed, and converted energy as a function of \co{} price. The 5\% target configuration is marked with a red circle in the top and middle plots.}
	\label{fig:target configuration}
\end{figure}

\begin{figure*}[!b]
	\centering		
	\textbf{5\%, 380~\euro/t\co{} \hspace{5.cm} 5\%, 280~\euro/t\co{}}
	\includegraphics[trim=0 0cm 0 0cm,width=0.49\linewidth,clip=true]{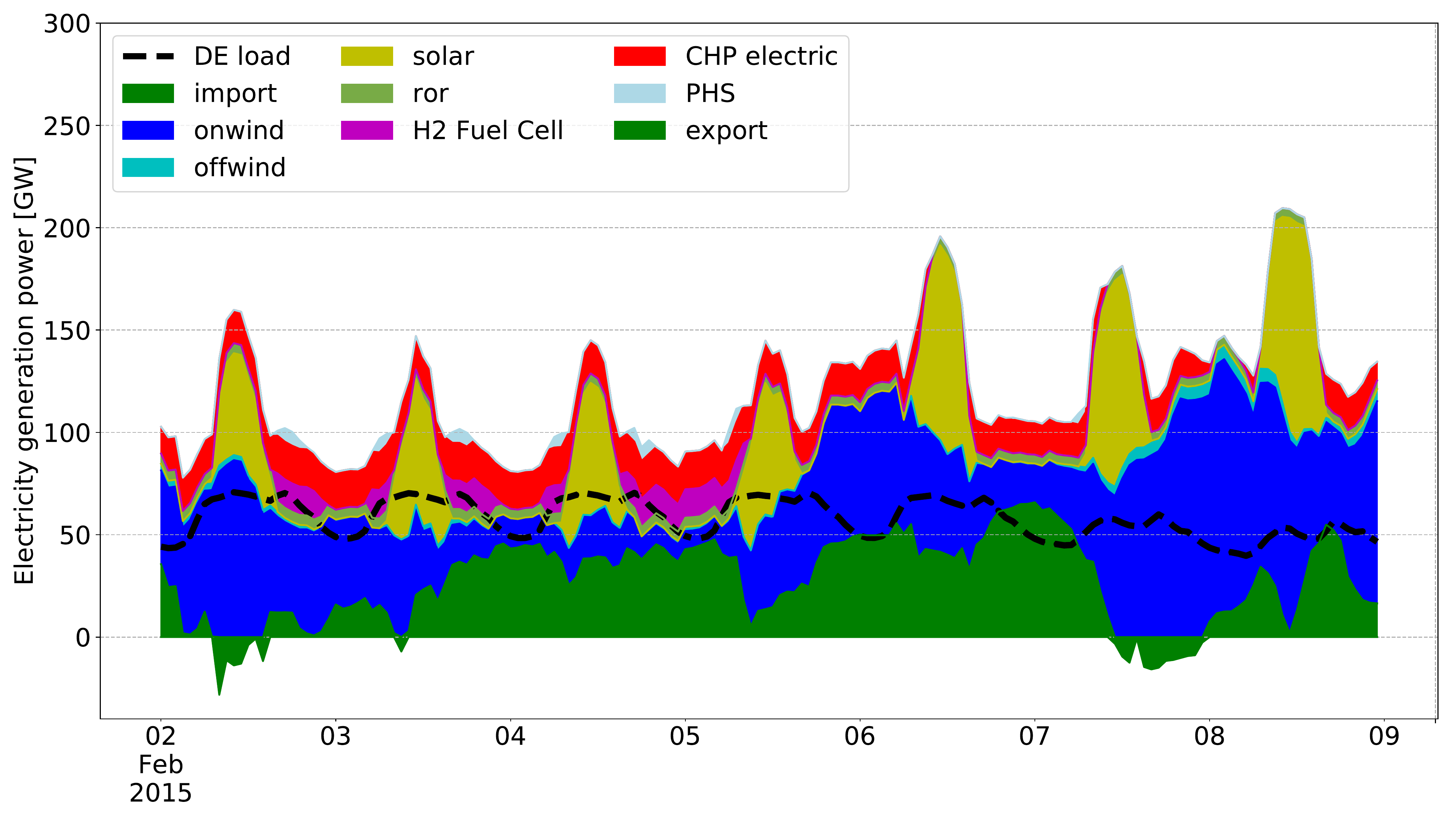}
	\includegraphics[trim=0 0cm 0 0cm,width=0.49\linewidth,clip=true]{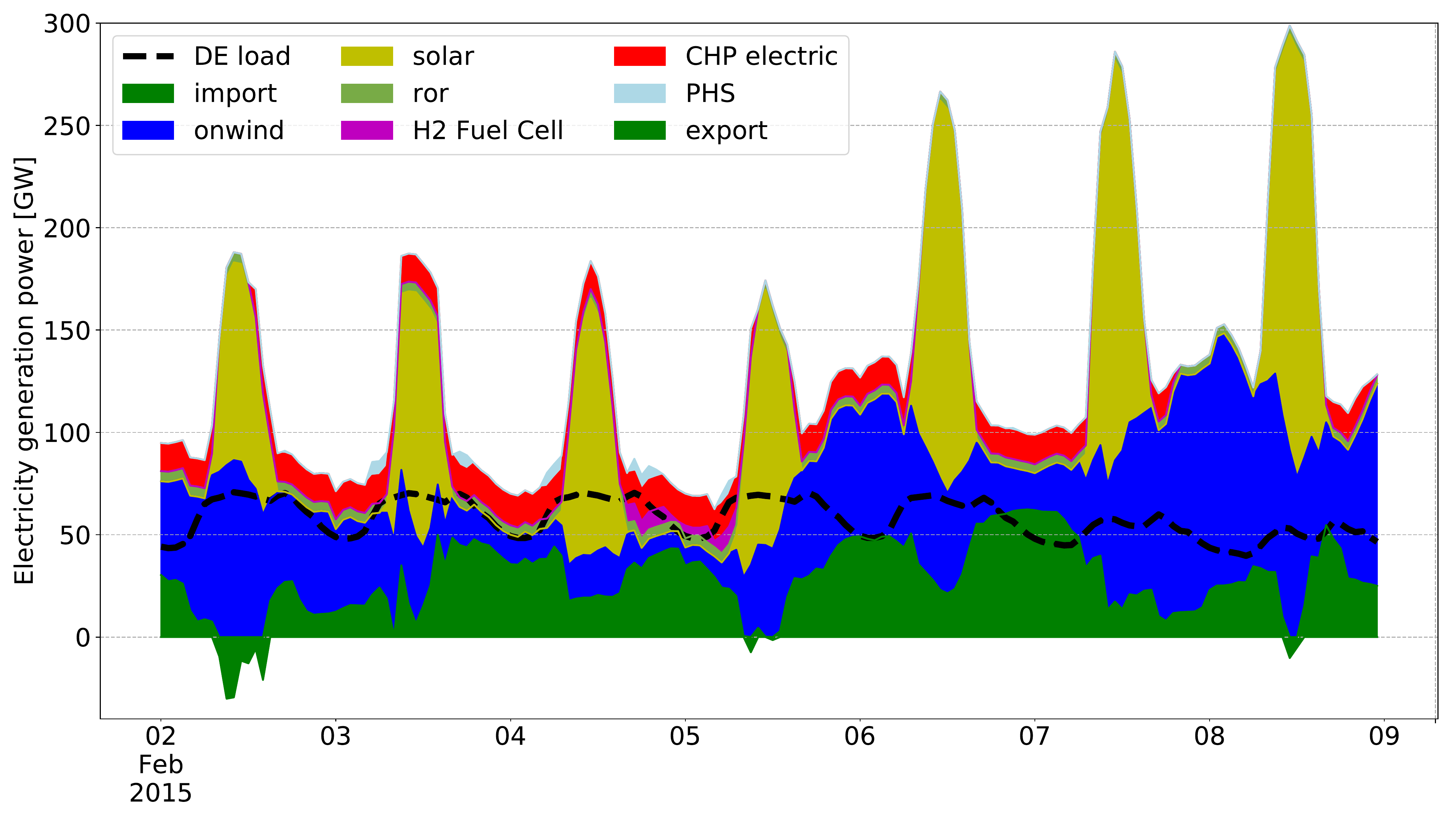}
	\includegraphics[trim=0 0cm 0 0cm,width=0.49\linewidth,clip=true]{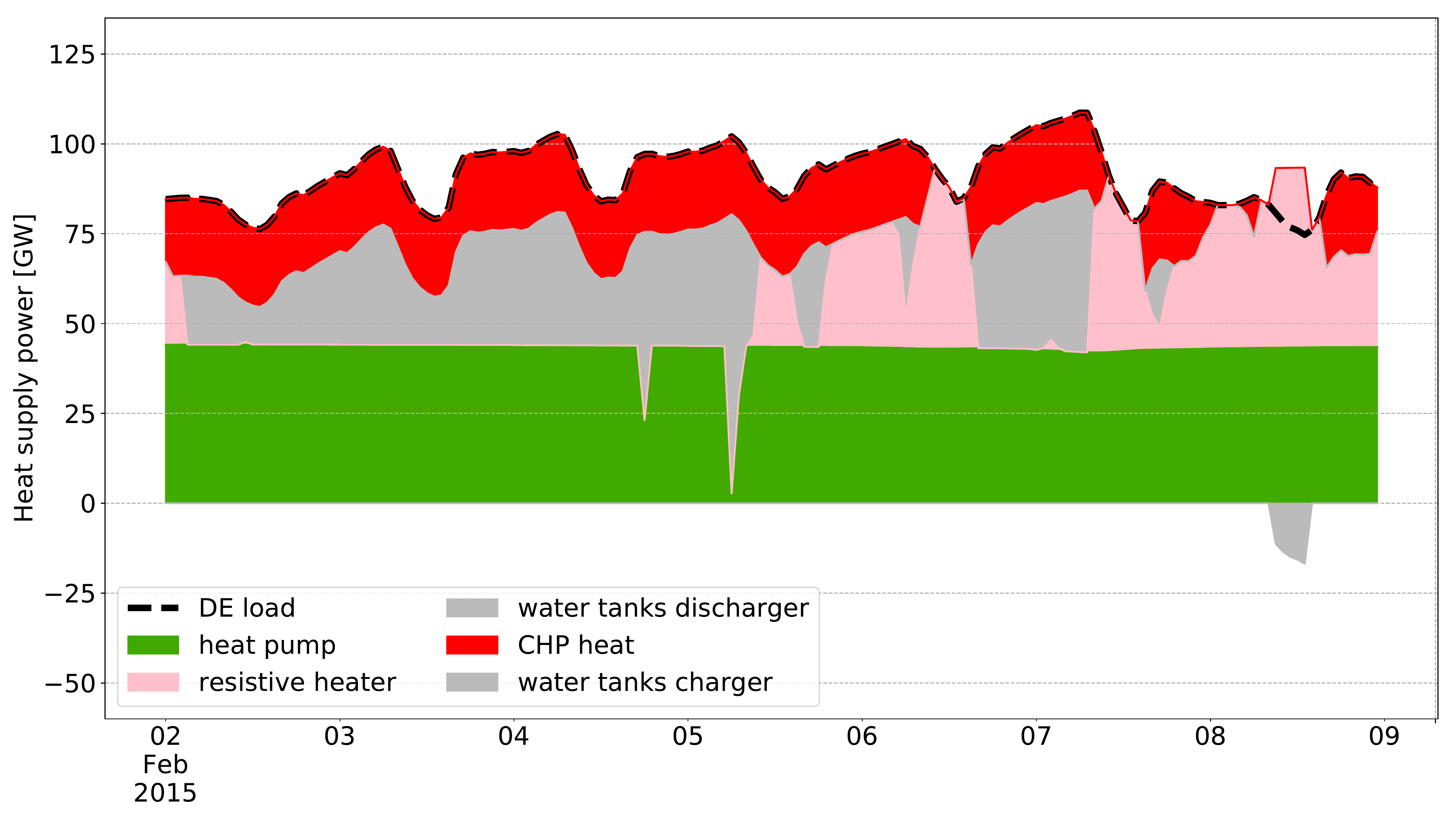}
	\includegraphics[trim=0 0cm 0 0cm,width=0.49\linewidth,clip=true]{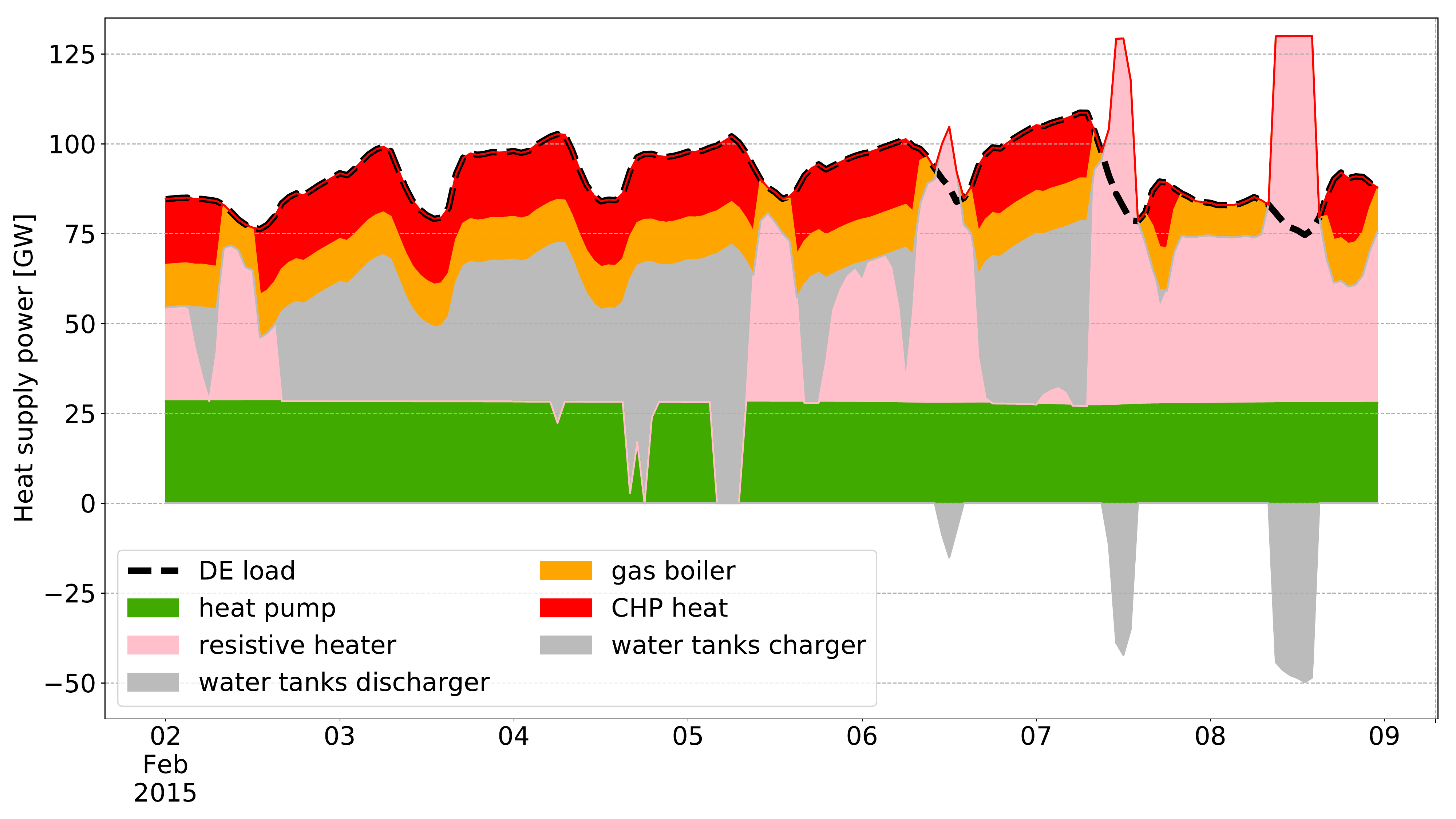}
	\caption{Electricity supply (top) and heating supply in densely-population areas (bottom) for the target configuration (left, 380~\euro/t\co{} with penetration of 0.64, which corresponds to the red circle in Figure \ref{fig:target configuration}) and the off-target configuration (right, 280~\euro/t\co{} with penetration of 0.9, which corresponds to the green circle in Figure \ref{fig:target configuration}) during a cold week in Germany.}
	\label{fig:time series}
\end{figure*}

\subsection{Target configuration}\label{subsec:target}
Although a sweep of VRES penetrations has been carried out in the previous section, it is our interest to investigate highly renewable energy systems in this paper. We do not focus on a particular future year but we analyse the cost-optimal system configurations which could fulfil the 5\%/10\%/20\% \co{} emissions, compared to 1990 levels.

As it was shown in Figure \ref{fig:sweep}, multiple combinations of gross VRES penetrations and \co{} prices are able to achieve the 5\% \co{} emission target. The questions that need to be addressed now are what is the cost-optimal \co{} price and how flat is that optimum. The top plot of Figure \ref{fig:target configuration} shows the \co{} emissions as a function of gross VRES penetration, assuming cost-optimal wind/solar mix, for various \co{} prices, ranging from 200~\euro/t\co{} to 500~\euro/t\co{}. For every \co{} price, the system LCOE intersecting the 5\% line is plotted in the middle of Figure \ref{fig:target configuration}, where solid/dashed line represents LCOE including/excluding \co{} taxes. The minimum of the latter curve determines the cost-optimal \co{} price, since, as previously mentioned, the money collected through \co{} tax could be reused in the system. The bottom plot of Figure \ref{fig:target configuration} shows the composition of generated and converted energy, as well as the curtailed renewable energy, as a function of \co{} price.

As \co{} price rises from 200~\euro/t\co{} to 320~\euro/t\co{}, the LCOE, either including or excluding tax associated with pricing \co{}, decreases rapidly. The required VRES penetration is significantly reduced, mainly due to two reasons: less curtailment and technologies with higher efficiencies are selected. The bottom of Figure \ref{fig:target configuration} depicts how curtailment declines from more than 2000 TWh\el/a at 200~\euro/t\co{} to almost 0 at high \co{} prices. Meanwhile, energy converted through heat pumps increases substantially from 0 to supply more than 3000 TWh\th/a and that converted by resistive heaters drops from 4000 TWh\th/a to less than 500 TWh\th/a. A system with higher \co{} price curtails less renewable energy and utilises better VRES generations, hence requires less VRES infrastructure and becomes cost-effective. LCOE excluding \co{} tax reaches the minimum at 380~\euro/t\co{}, and then starts to increase slowly due to the fact that expensive technologies, such as batteries and heat pumps, are extensively deployed. The even higher \co{} prices incentivise a more efficient VRES-dominated but slightly more expensive system. As a result, the cost-optimal \co{} price to achieve the 5\% emission target has been identified as 380~\euro/t\co{} (marked with a red circle in Figure \ref{fig:target configuration}). The corresponding system configuration is referred to as the 5\% target configuration, where gross VRES penetration is 0.64 and gross wind/solar mix is 0.8. The LCOE excluding \co{} tax shows low sensitivity around the optimum. \co{} prices ranging from 320~\euro/t\co{} to 500~\euro/t\co{} modify the LCOE by less than 1~\euro/MWh. Furthermore, the compositions of the generated and converted energy are roughly flat around the target configuration. 

In the same manner, the \co{} prices required to achieve 10\% and 20\% \co{} emissions have been identified to be 260 and 160~\euro/t\co{} respectively, and the corresponding gross VRES penetrations are 0.57 and 0.46; see Table \ref{tab:configuration}. At this point, it needs to be mentioned that the optimal \co{} price is very sensitive to the technology mix. In fact, the \co{} price forces the choice of expensive heat supply from power-to-heat (PTH) technologies, \textit{i.e.}, heat pumps over using cheap gas by altering the merit order. The cost-optimal \co{} price identified in this work are impacted by the fact that our current model excludes biomass as a possible technology to supply electricity and heat. Incorporating this technology is expected to reduce the \co{} price necessary to alter the merit order and force the gas out of the system. 

Figure \ref{fig:time series} depicts the electricity and heating supply for Germany during the first week of February, comparing two system configurations which are both able to achieve 5\% \co{} emissions. The left panel shows the target configuration, where the \co{} price is 380~\euro/t\co{} and VRES penetration is 0.64 (red circle in Figure \ref{fig:target configuration}). The right panel presents the so called `off-target' configuration, where \co{} is 280~\euro/t\co{} and VRES penetration is 0.9 (green circle in Figure \ref{fig:target configuration}). Hourly electricity generation fluctuates less for the target configuration, due to lower solar penetration. During the hours of low VRES generation, Germany imports considerable amount of electricity, which underlines the importance of transmission for highly renewable system. The surplus electricity generation is either stored or converted into heat. The target case relies heavily on efficient heat pumps, supplying nearly half of the heating demand. Resistive heaters are only activated if there is any electricity surplus left after the heat pumps, such as the last two days of this cold week. Finally, water tanks discharge and CHP plants supply the remaining heating demand. Heat pumps play a less powerful role for the off-target case, supplying less than one third of the thermal load. Resistive heaters take advantage of the surplus electricity, even being able to charge water tanks for certain hours. Less efficient technologies are still extensively used in the off-target case, and imply a more expensive system. In both configurations, heat pumps operates in a more stable manner compared to resistive heaters and water tanks.

\subsection{Spatial distribution} \label{subsec:distribution}

\begin{figure*}[!t]
	\centering		
	\textbf{System cost \hspace{4.65cm} Primary energy}
	\includegraphics[trim=0 0cm 0 0cm,width=0.4\linewidth,clip=true]{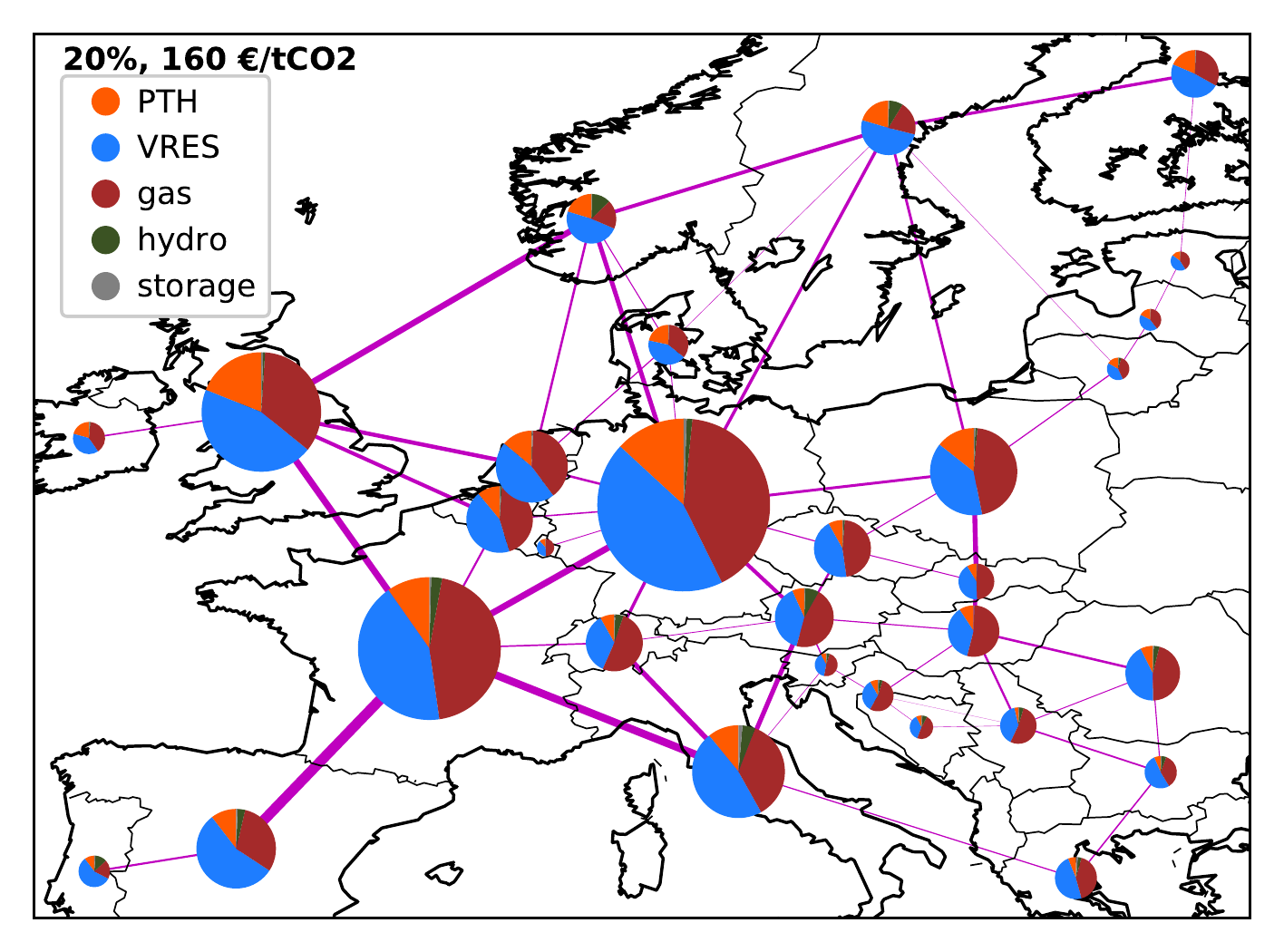}
	\includegraphics[trim=0 0cm 0 0cm,width=0.4\linewidth,clip=true]{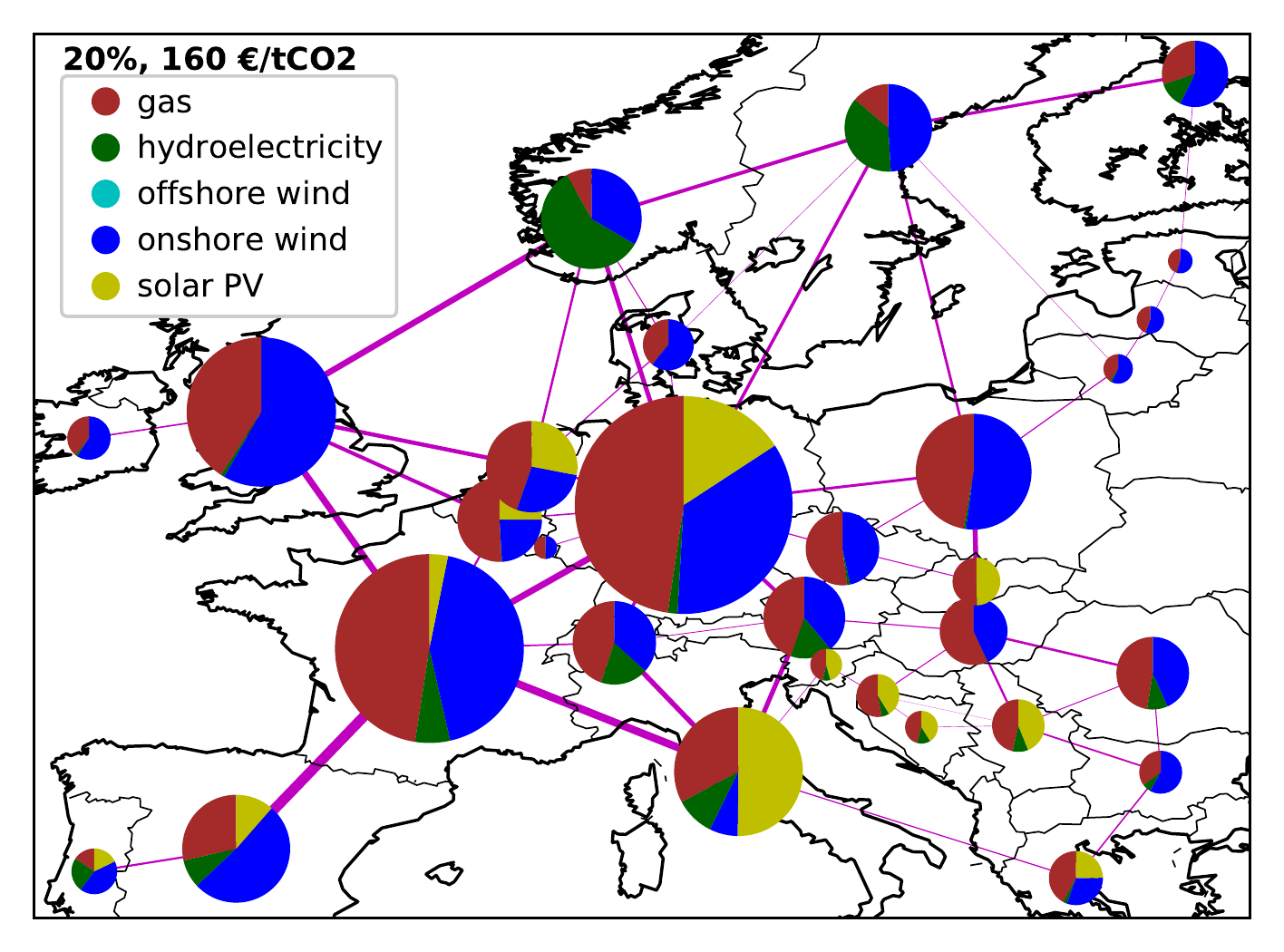}
	\includegraphics[trim=0 0cm 0 0cm,width=0.4\linewidth,clip=true]{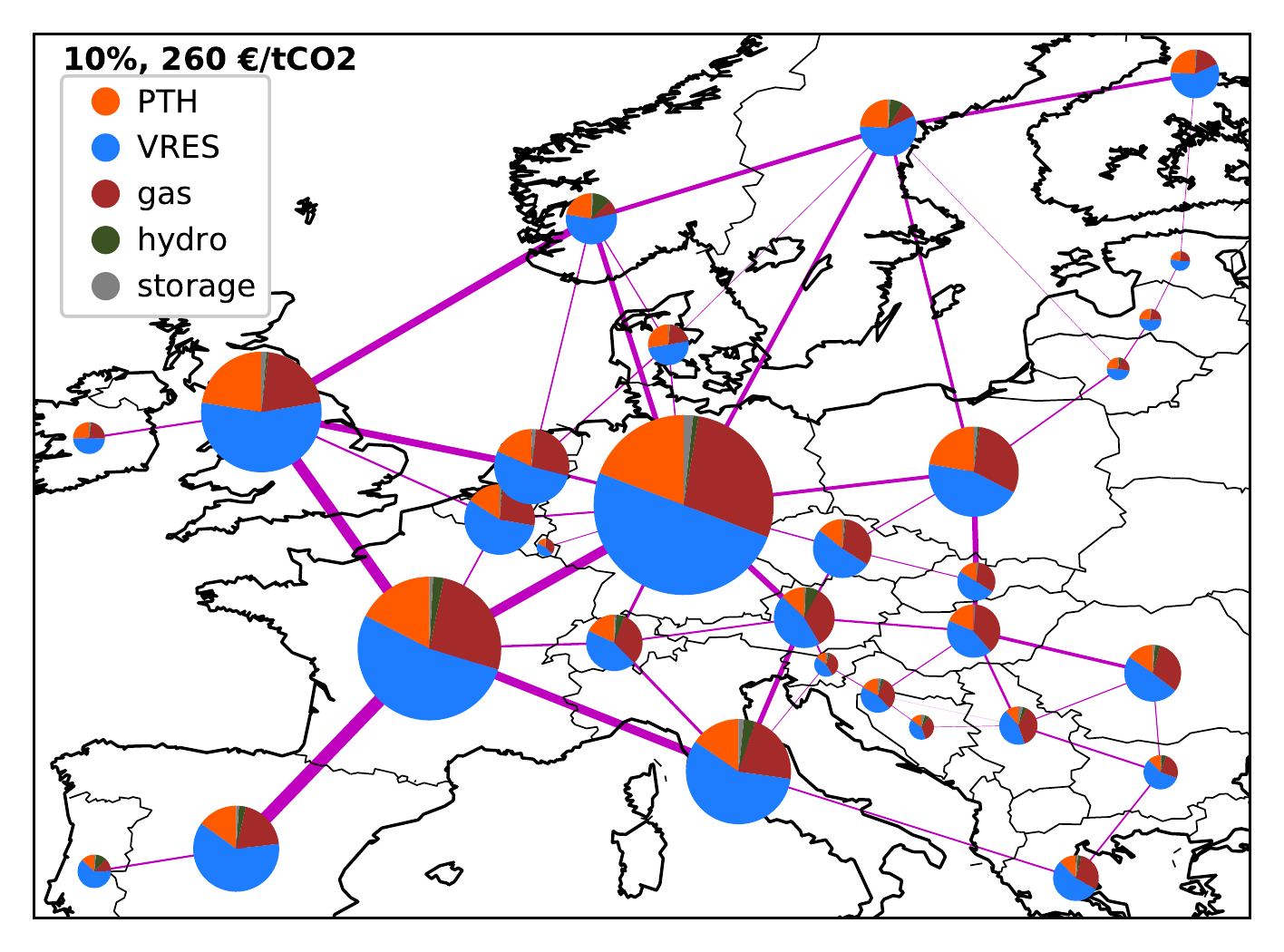}
	\includegraphics[trim=0 0cm 0 0cm,width=0.4\linewidth,clip=true]{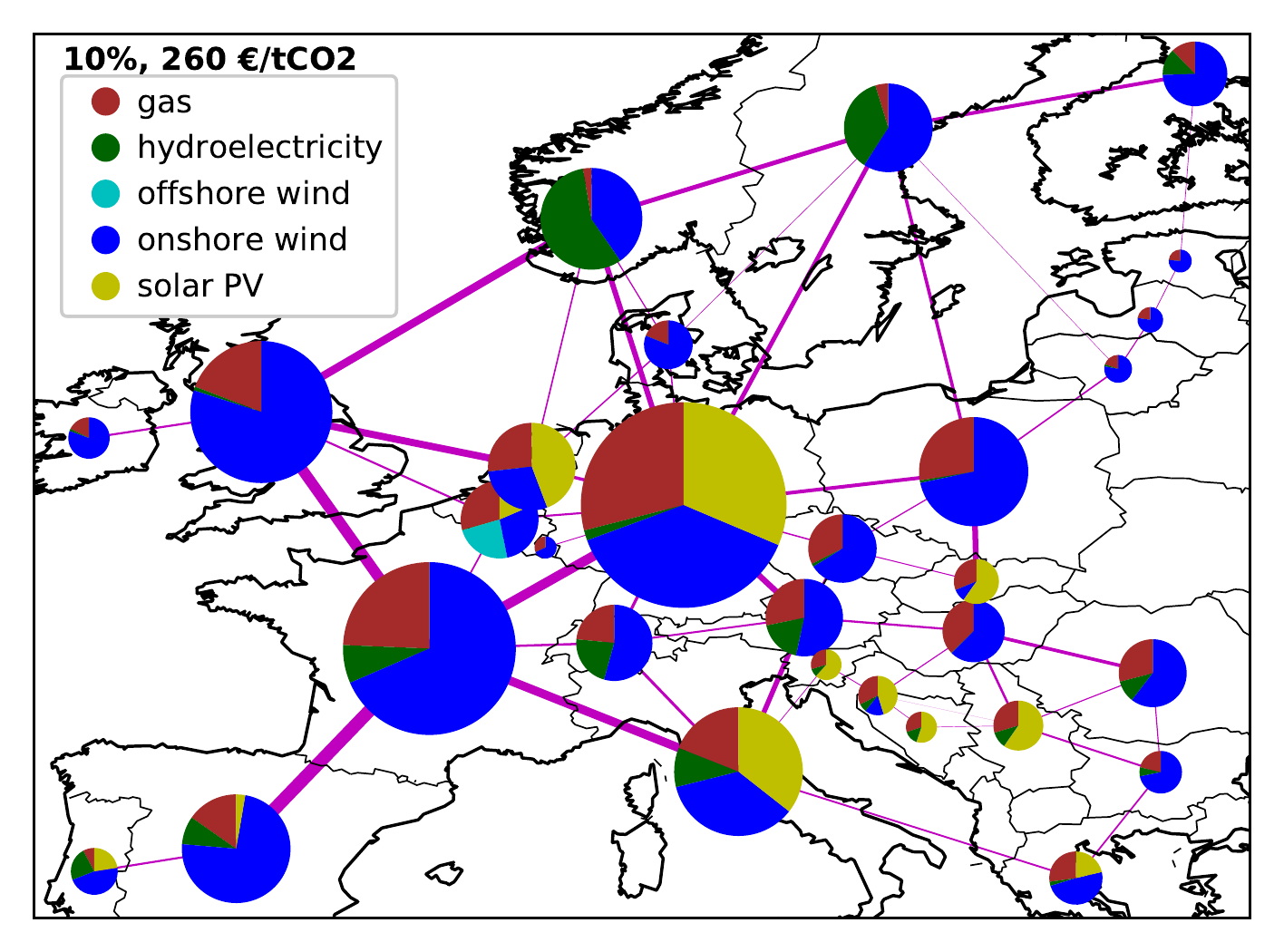}
	\includegraphics[trim=0 0cm 0 0cm,width=0.4\linewidth,clip=true]{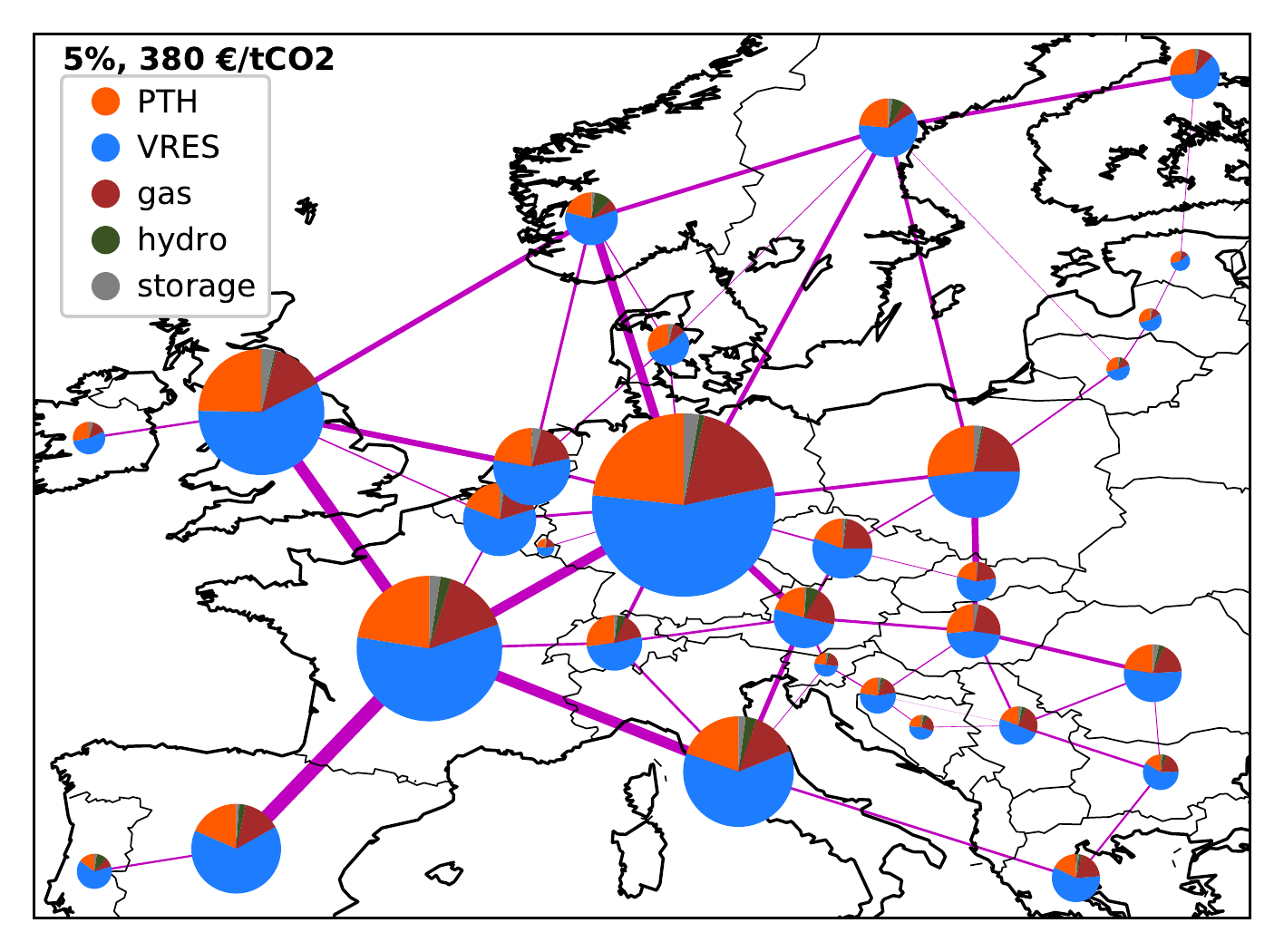}
	\includegraphics[trim=0 0cm 0 0cm,width=0.4\linewidth,clip=true]{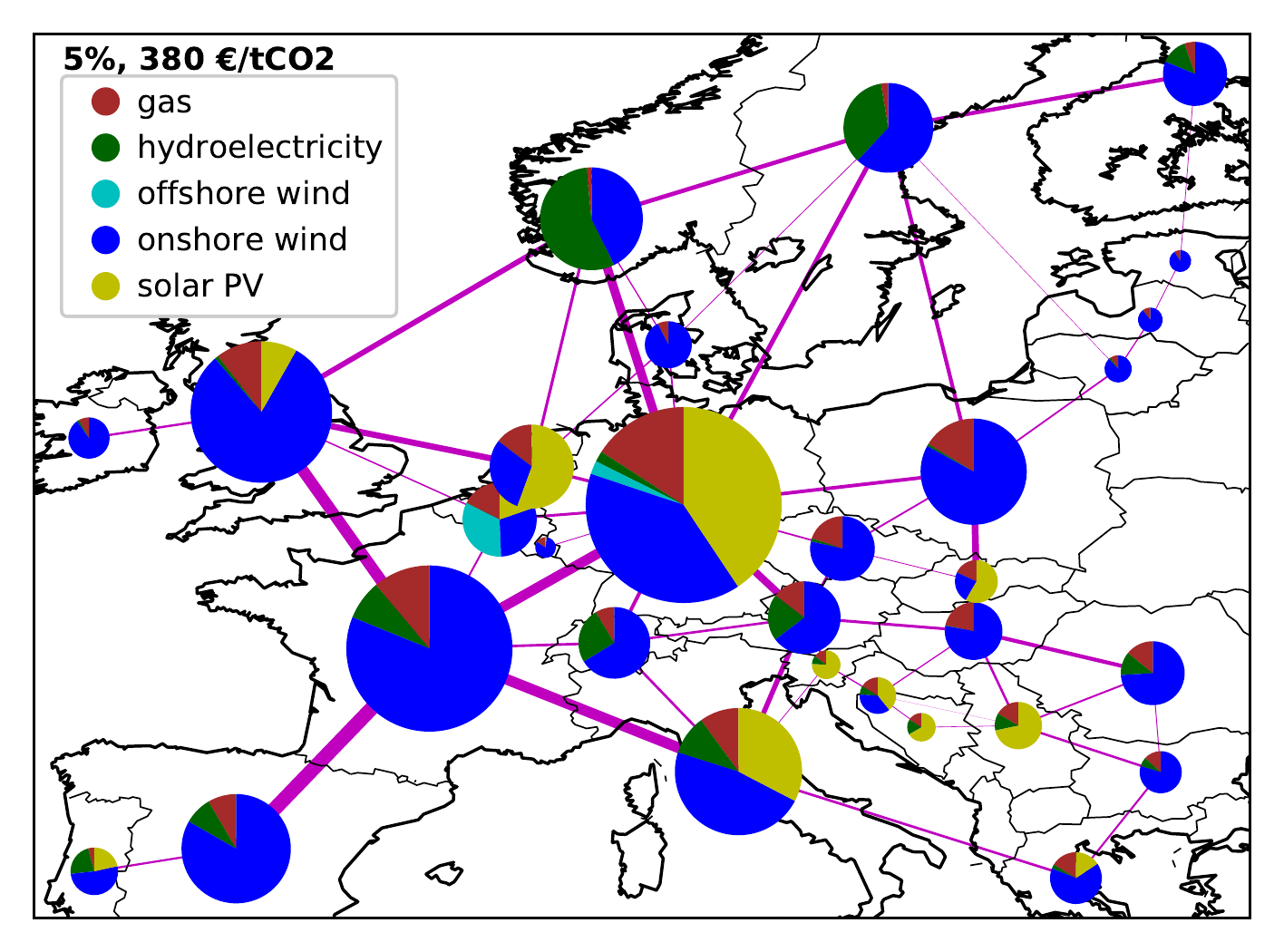}
	\caption{Spatial distributions of total system cost (left) and primary energy (right) for 20\%, 10\%, and 5\% \co{} emission target configurations (from top to bottom). System cost consists of VRES (including solar PV, onshore, and offshore wind), PTH (including heat pumps and resistive heaters), gas (including \co{} tax), storage (including hydrogen, battery, and hot water tank storage) and hydroelectricity (whose capacities are fixed). Primary energy consists of onshore wind, solar PV, offshore wind, gas, and hydroelectricity.}
	\label{fig:spatial layouts}
\end{figure*}

The main characteristics of the 20\%, 10\%, and 5\% target configurations are analysed in this subsection. The left column of Figure \ref{fig:spatial layouts} depicts the spatial distributions of total system costs. The system costs are dominated by VRES and PTH technologies in most countries, and the share of the latter increases as more ambitious \co{} objectives are assumed. The total system costs excluding \co{} tax for the 20\%, 10\%, and 5\% \co{} target are 277, 320, and 355 billion~\euro/a, respectively. The increments mostly come from VRES costs which are 142, 182, and 207 billion~\euro/a and PTH technologies costs which represent 40, 64 and 83 billion~\euro/a respectively. As higher \co{} prices are imposed to the system, more VRES and PTH capacities are installed, moving the system configuration towards a more efficient and low-carbon electricity-and-heating coupled system.

This is also shown in the right column of Figure \ref{fig:spatial layouts}, where the spatial distributions of primary energy are presented. For the 20\% target configuration (160~\euro/t\co{}), primary energy from gas is comparable to VRES in most countries. As the allowed \co{} emissions are reduced, the total amount of primary energy which has been produced also declines. This can be observed in Figure \ref{fig:spatial layouts}, where the size of the pie chart in every country represents the total primary energy consumption. The major reduction of primary energy comes from less usage of gas, which is encouraged by deploying and exploiting VRES more effectively. %The spatial distribution of primary energy shows the splitting pattern of resource utilisation among different regions in Europe, enabling the best VRES potentials to be exploited in every country. Similarly to previous results \cite{Schlachtberger_2017,Brown_2018}, northern countries exploit wind resource while southern countries prefer to install more PV. 

\begin{figure*}[!t]
	\centering
	\textbf{Thermal energy \hspace{3.7cm} Thermal capacity}
	\includegraphics[trim=0 0cm 0 0cm,width=0.4\linewidth,clip=true]{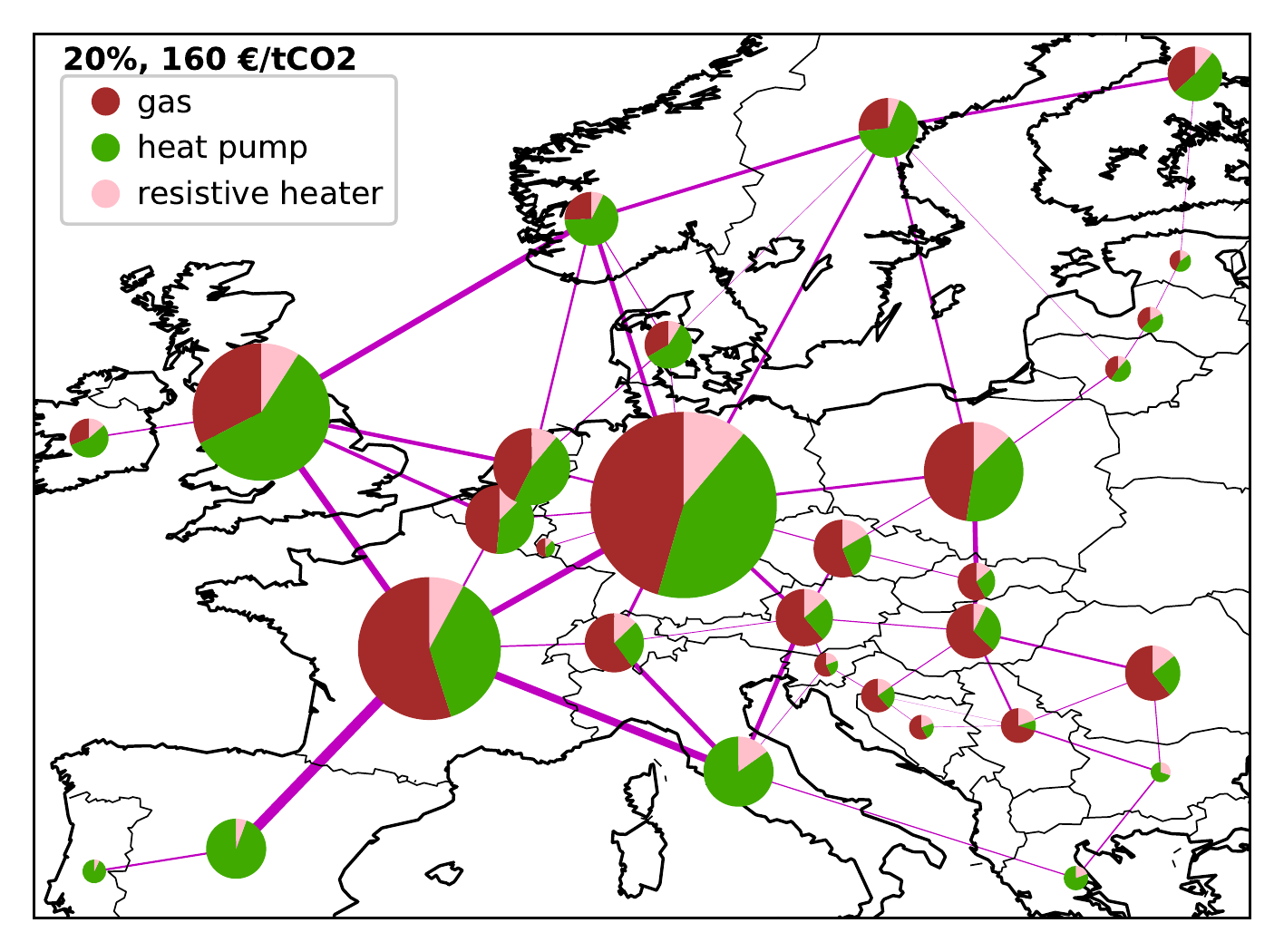}
	\includegraphics[trim=0 0cm 0 0cm,width=0.4\linewidth,clip=true]{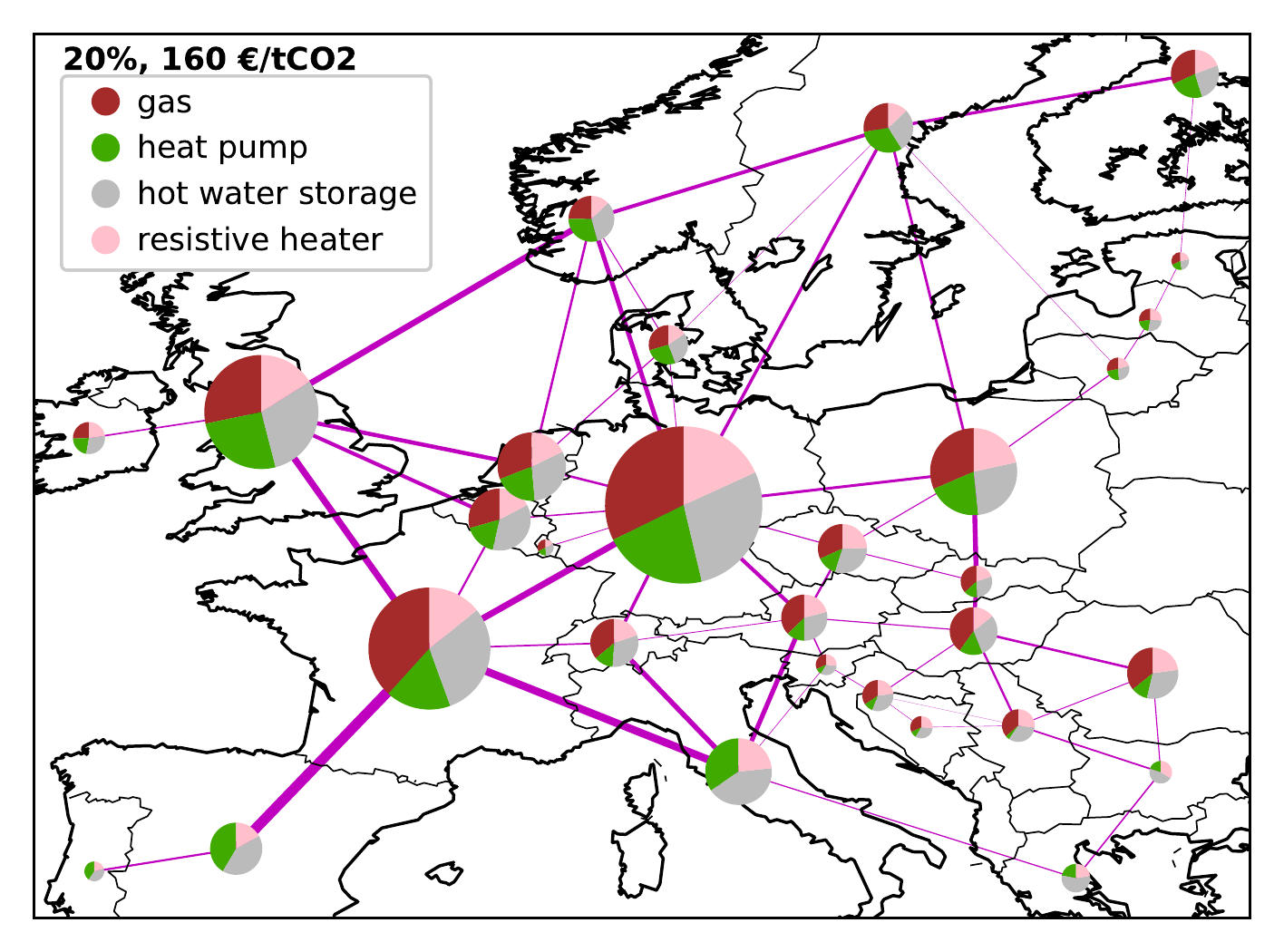}
	\includegraphics[trim=0 0cm 0 0cm,width=0.4\linewidth,clip=true]{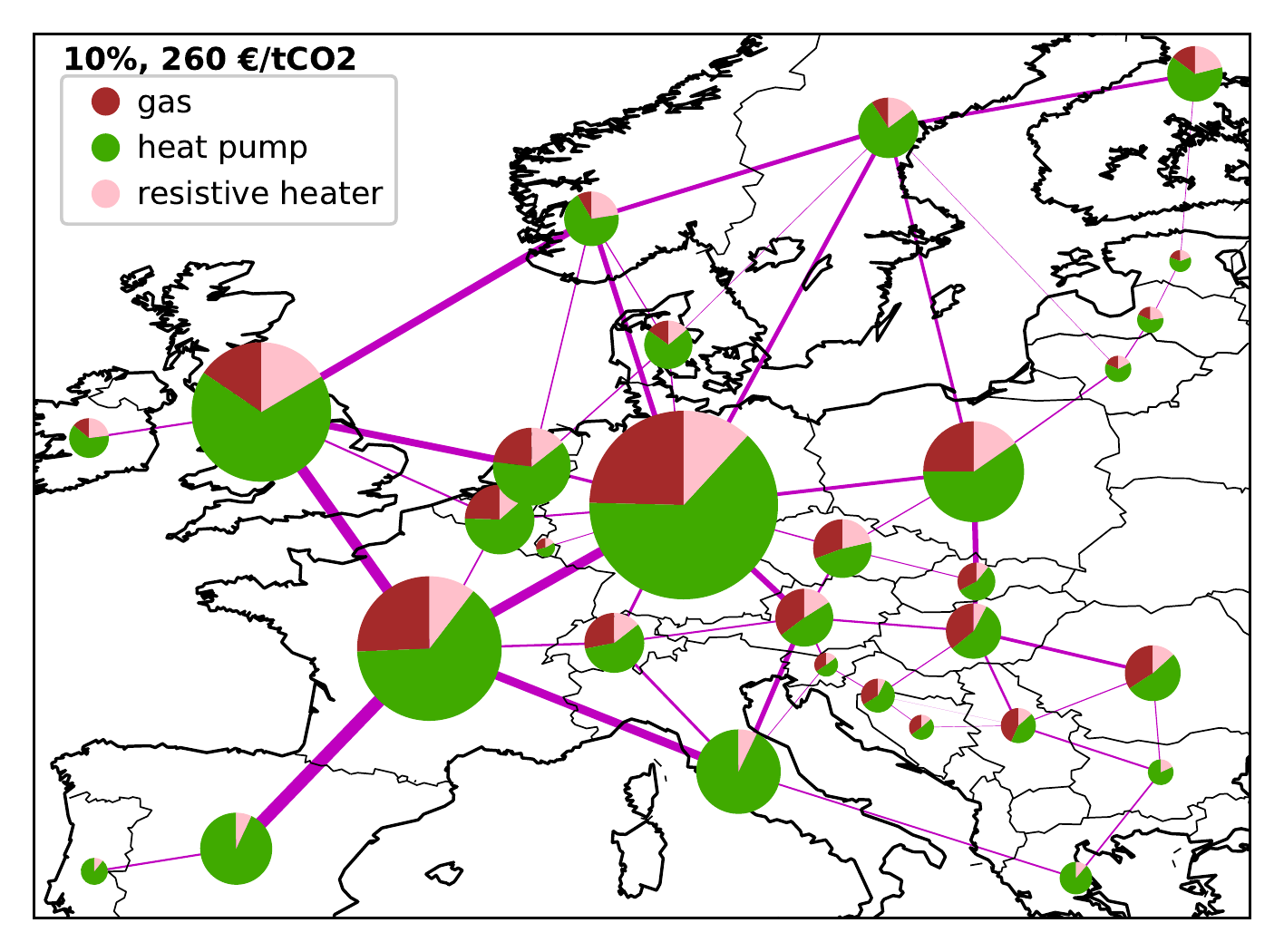}
	\includegraphics[trim=0 0cm 0 0cm,width=0.4\linewidth,clip=true]{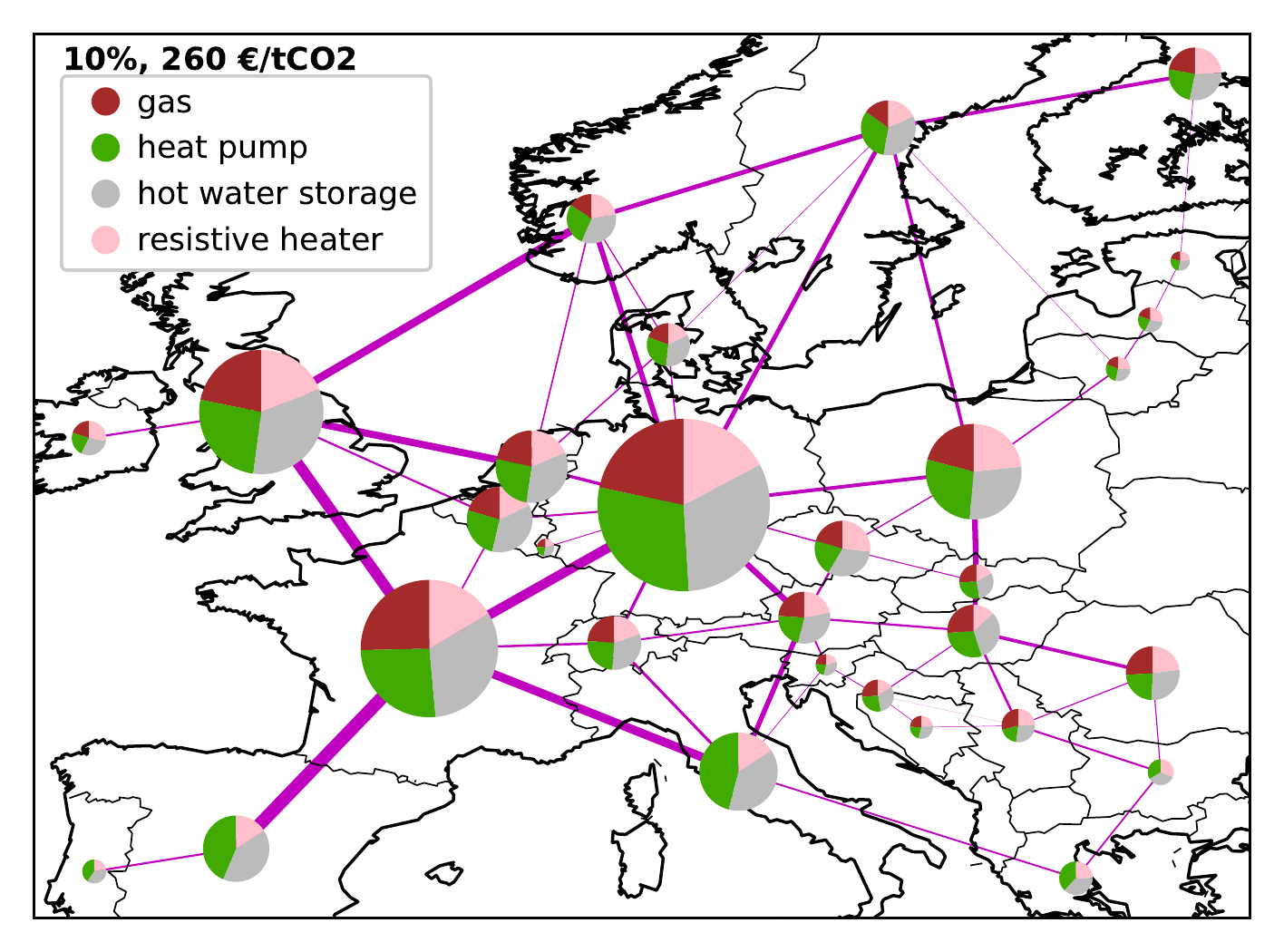}
	\includegraphics[trim=0 0cm 0 0cm,width=0.4\linewidth,clip=true]{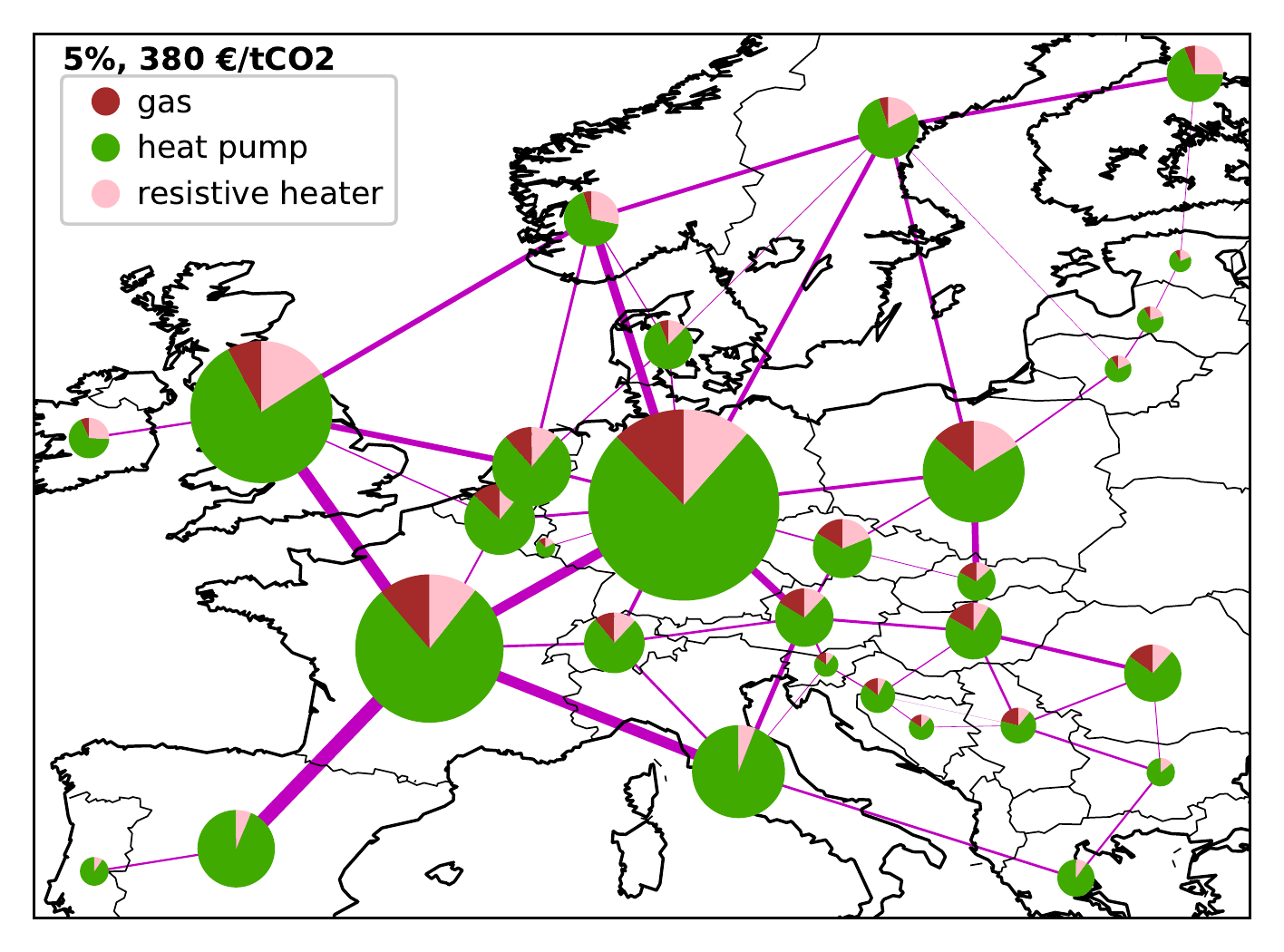}
	\includegraphics[trim=0 0cm 0 0cm,width=0.4\linewidth,clip=true]{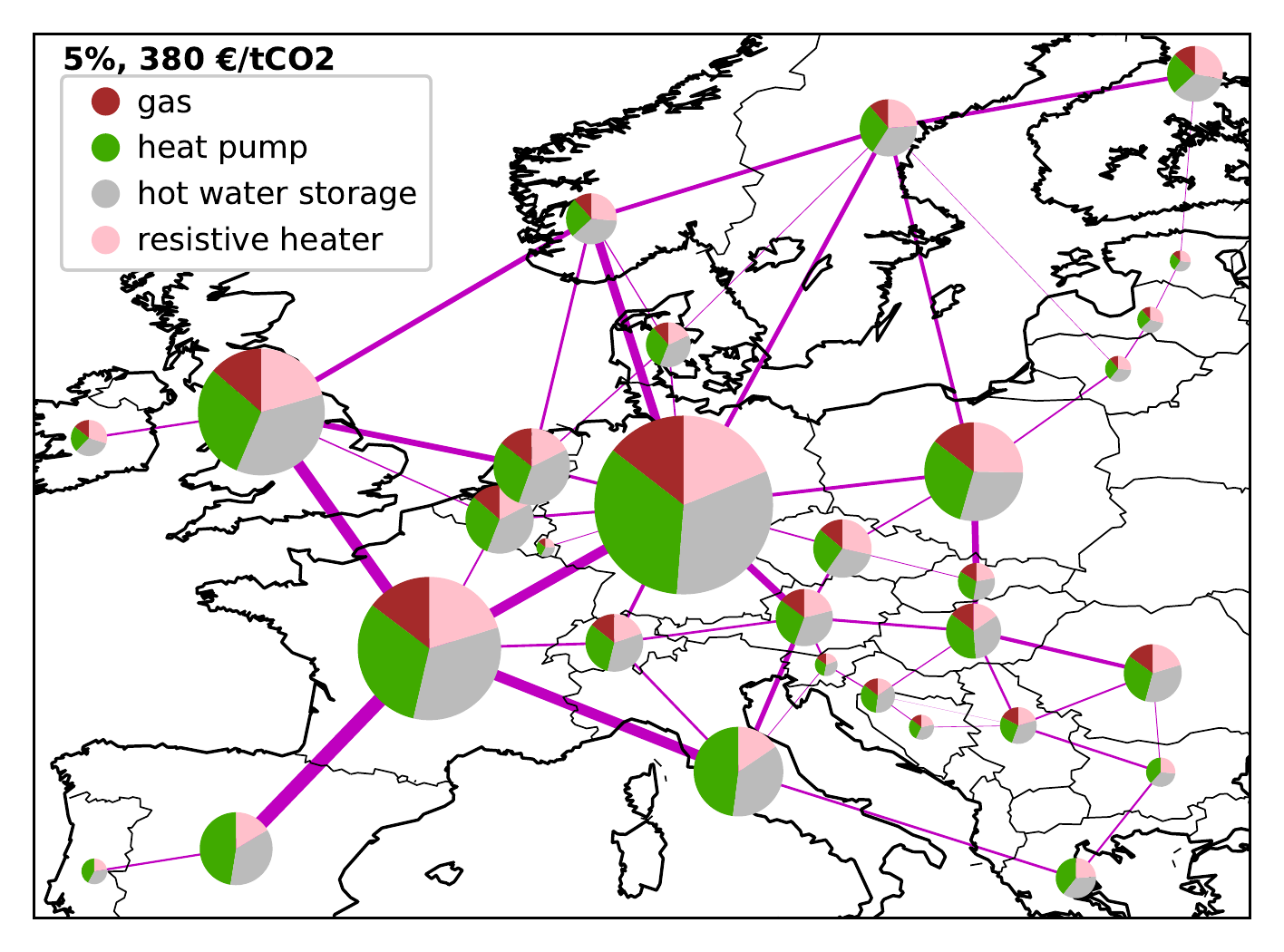}
	\caption{Spatial distribution of thermal energy (left) and thermal capacities (right) for 20\%, 10\% and 5\% \co{} emissions target configurations (from top to bottom). The thermal energy is calculated as the generated energy in the urban and rural heat buses, while thermal capacities are calculated as the capacities installed in those buses.}
	\label{fig:spatial layouts heat}
\end{figure*}

As \co{} emissions decline, gas is replaced by onshore wind in most of the countries, even in Italy and Spain where solar PV has lower LCOE compared to onshore wind. Although solar PV itself is cheaper than onshore wind in those countries, it requires higher amount of storage, typically electric batteries, in order to balance its diurnal variation. Consequently, most of the countries prefer to install more wind than solar, as transmission lines are optimised freely. If the transmission lined are constrained to a lower level, such as todays volume, the share of solar PV increases significantly in countries like France and Spain. In a few countries solar substitutes gas to become the dominant resource, such as in Germany and the Netherlands. Even though the onshore wind resources in those countries turn out to be better than solar PV, the geographical limits (Equation \ref{eq:geo limit}) for onshore wind have been reached, hence solar PV becomes the next choice. 

\begin{table*}[!t]
\centering
\begin{threeparttable}	
\caption{Aggregated system configurations\tnote{a}}\label{tab:configuration}
\begin{tabular}{l|ccc|ccc}
\toprule
Transmission volume\tnote{b} & \multicolumn{3}{c}{Optimal volume} & \multicolumn{3}{c}{Todays volume} \\
Emission level&     20\% &    10\% &     5\% &    20\% &    10\% &      5\% \\
\midrule
\co{} price                 &     160 &    260 &    380 &    200 &    320 &     580 \\
Gross penetration           &    0.46 &   0.57 &   0.64 &    0.5 &   0.64 &     0.7 \\
Gross wind/solar 			&    0.77 &    0.8 &    0.8 &   0.73 &   0.74 &    0.79 \\
System cost incl. \co{} tax &     348 &    378 &    397 &    380 &    417 &     456 \\
System cost excl. \co{} tax &     277 &    320 &    355 &    291 &    346 &     391 \\
LCOE incl. \co{} tax        &   54.3 &  58.9 &  61.9 &  59.2 &  64.9 &   71.1 \\
LCOE excl. \co{} tax        &   43.2 &  49.8 &  55.4 &  45.4 &  53.9 &   60.9 \\
Onshore wind                &    1,090 &   1,406 &   1,567 &   1,126 &   1,428 &    1,591 \\
Offshore wind               &       0 &     10 &     21 &      5 &     33 &      88 \\
Solar PV                    &     542 &    616 &    719 &    703 &    902 &     812 \\
Resistive heater            &     307 &    389 &    464 &    434 &    581 &     673 \\
Heat pump                   &      69 &    113 &    148 &     67 &    103 &     143 \\
Gas boiler                  &     567 &    469 &    332 &    512 &    399 &     300 \\
OCGT                        &       0 &      0 &      0 &     17 &      1 &       0 \\
CHP                         &     363 &    243 &    165 &    464 &    336 &     268 \\
Battery storage             &       9 &     10 &      0 &    145 &    180 &     143 \\
Hydrogen storage            &       0 &      0 &      0 &      0 &      0 &       0 \\
Hot water tank              &    7,768 &  27,823 &  91,796 &  17,232 &  57,818 &  156,753 \\
Transmission volume         &     141 &    176 &    196 &     32 &     32 &      32 \\
\bottomrule
\end{tabular}
\begin{tablenotes}
\footnotesize
\item [a] \co{} price, system costs and LCOE are in units of \euro/t\co{}, billion~\euro/a and \euro/MWh, respectively. Generators and converters are in terms of power capacities (GW), storage units are in terms of energy capacities (GWh), and transmission volume is in terms of power capacities multiplied by their lengths (TWkm).
\item [b] Todays transmission volume is the sum of todays Net Transfer Capacities (NTCs, 31 TWkm) and the under-construction lines (assumed to be 0.6 TWkm). 
\end{tablenotes}
\end{threeparttable}
\end{table*}

The spatial distribution of produced heat and capacities of heating technologies are shown in Figure \ref{fig:spatial layouts heat}. For the 20\% target configuration (160~\euro/t\co{}), gas is the main contributor to the generated thermal energy. Resistive heaters play an important role in converting electricity to heat. As the \co{} emissions target declines to 10\% (260~\euro/t\co{}), heat pumps substitute gas to become the main heating technology, and the shares of resistive heaters and hot water tanks in terms of thermal capacities decrease to a smaller extent. This is due to the fact that, while heat pumps provide the base demand, resistive heaters and hot water tanks cover the peak heating demand, as it is shown in Figure \ref{fig:time series}. Another proof of these behaviours is the full power capacity operating time, computed by dividing the supplied thermal energy by the thermal capacity. For the 5\% target configuration (380~\euro/t\co{}), this number represents 4,320, 1,164 and 1,491 hours per year for heat pumps, resistive heaters, and gas boilers respectively. Gas still represents a non-negligible share of the thermal capacities in the 5\% target configuration (380~\euro/t\co{}).

Table \ref{tab:configuration} lists the aggregated system configurations for all the technologies involved in the model. Under the assumption of optimal transmission, the onshore wind plays a more pivotal role as \co{} emissions decline, hence incentivising a strongly connected European transmission system. All the heating technologies see a decrease as the system decarbonises, except for PTH and hot water tanks, whose combination can balance the variations of VRES and demand. The need for electric batteries is also lower, since PTH technologies in well-coupled systems could utilise the excess electricity generations directly instead of storing. The hot water tanks expand its energy capacity tremendously from the 20\% to 5\% target scenario, while still remaining less than 2\% of the total system cost. The cheap, large hot water tanks with a time constant of 180 days enable heating demand to be shifted seasonally.

For the 5\% target configuration (380~\euro/t\co{}), the optimal transmission volume is 196 TWkm which is roughly 6 times higher than todays NTCs (31.6 TWkm). Such a large grid extension might be infeasible due to social acceptance issues. The right three columns in Table \ref{tab:configuration} restrict the transmission volume to todays value. The capacity of the links do not exactly correspond to current NTCs, but the CAP limit, defined in Equation (\ref{eq:lvcap}), corresponds to todays transmission volume. Note that the cost-optimal \co{} prices as well as the corresponding gross VRES penetrations increase to a certain extent. Without a strongly-connected transmission system, each country has to rely on its own to a large extent, especially on storage, and the benefits of integrating high shares of VRES are narrowed down. At the same time, the needs for building up other infrastructures are pronounced, such as PTH and CHP. In particular, the storage requirements show a substantial expansion. For instance, in order to achieve the 5\% \co{} emissions target, todays transmission volume demands approximately extra 143 GWh of battery and 65,000 GWh of water tank compared to optimal transmission case. This clearly emphasises the crucial role played by the transmission system for highly renewable energy systems.

\section{Discussions} \label{sec:discussion}

\subsection{Comparison to similar studies}
%TOTAL system cost
For the 5\% target configuration (380~\euro/t\co{}), the total system cost excluding \co{} tax is found to be 355 billion~\euro/a, while the LCOE excluding \co{} tax is 55.4~\euro/MWh respectively. Brown \textit{et al.} \cite{Brown_2018} examined the synergies of sector coupling under similar cost assumptions. Their analysis included electricity, heating, and transportation, where the demand in the transportation sector was represented by electric vehicles, whose number and cost were exogenous to the model. Brown \textit{et al.} estimated a total system cost equal to 440 billion~\euro/a and 57.5~\euro/MWh for the LCOE excluding \co{} tax. The difference of total system cost is mainly due to two reasons. On one hand, transportation sector is also included in Brown \textit{et al.}, which consumes an additional electricity demand of 1,102 TWh\el/a, thus driving up the total cost. On the other hand, more flexibility is offered and less constraints are binding in Brown \textit{et al.}, thus bringing down the total cost. The authors also considered solar thermal as direct heat-supply technology and the possibility of using methanation to transform excess electricity into gas. Additional constraints are imposed in our work, \textit{i.e.}, the \WH layout restricts locations with high VRES potentials to be utilised, which might result in a less cost-competitive system. In terms of LCOE excluding \co{} tax, the two results are very close. The difference of 2.1~\euro/MWh mainly lies in the fact that producing electricity for the transportation sector is generally more expensive than producing heat.

Schlachtberger \textit{et al.} \cite{Schlachtberger_2017} investigated the benefits of transmission in a highly renewable European electricity network with similar costs assumptions. Under the 95\% \co{} emission reduction constraint, relative to 1990, and assuming optimal transmission, they found an LCOE of 64.8~\euro/MWh. This difference is most probably due to the fact that Schlachtberger \textit{et al.} did not consider heating demand in their model, where the heating sector is cheaper to be supplied compared to electricity.

%VRES mix
The 5\% target configuration (380~\euro/t\co{}) assuming optimal transmission implies gross VRES penetration equals to 0.64 and optimal gross wind/solar ratio of 0.8. Ashfaq \textit{et al.} \cite{Ashfaq_2017} studied the cost-minimised design of a highly VRES electricity-heating coupled network for regional, country and pan-European networks. The authors concluded that the excess hourly electricity generation when the annual VRES generation is equal to electricity demand is not enough to fully cover the heating sector. However, if the VRES energy generation is increased by 50\%, which corresponds to gross penetration of 0.67 in this paper, the required backup energy becomes minimal and the optimal wind/solar mix found was 0.7 when using heat pumps and 0.9 when using resistive heaters. Ashfaq \textit{et al.} also pointed out that a connected pan-European electrical grid with decentralised regional heat pump coupling seems promising, which our findings confirmed.

%TRANSMISSION
Transmission volume for the 5\% target configuration (380~\euro/t\co{}) in our model is 196 TWkm. Brown \textit{et al.} \cite{Brown_2018} found the optimal transmission volume to be 359 TWkm. Since those authors also integrated the transportation sector, more primary energy is produced by VRES, and consequently, a higher transmission volume is required. In addition, the \WH layout proposed in this paper limits the benefit of transmission to some extent, since all the countries are equally VRES self-sufficient.

\subsection{Limitations of the study}
The approach followed in this work entails certain limitations. Firstly, the sector coupling in this paper only includes electricity and heating sectors. Future works should further develop this analysis by including other sectors, such as transportation and industry. Although today's cooling demand is included in the electricity sector, a rise in cooling demand beyond today's levels is not considered. The cooling demand is expected to increase in the future, in particular as a result of climate change.

Secondly, there are other renewable sources or alternative technologies that will help to decarbonise the energy system that are not included in our model, such as solar thermal, concentrated solar power (CSP), biomass, waste heat, geothermal or nuclear. More flexibilities could be offered with richer choices of technologies, which might bring down the total system cost or optimal \co{} price. Gils \textit{et al.} \cite{Gils2017a} analysed the necessity of grid expansion and backup capacity as the share of VRES increases, including some of the technologies previously mentioned. The conclusion provided by them is that biomass, geothermal, and nuclear are not selected in the optimisation due to their high costs. For backup technologies, we only consider OCGT power plants, excluding coal or oil. Although a significant coal capacity is currently under operation in Europe, this technology is not expected to be part of the energy scenarios with high \co{} reductions in which this paper focuses.

Thirdly, only one single year of historical weather data (2015) has been modelled. Since the optimal wind/solar mix is very likely sensitive to the capacity factors of wind and solar, different weather years may lead to a slightly different picture \cite{Collins_2018}. For the sake of simplicity, electricity and heating demand are fixed to today's values and considered to be inelastic, which can be improved by implementing heat savings and demand-side management (DSM). Other factors that could raise the sensitivity problem are the cost assumptions, such as the ratio of capital costs between wind and solar. Schlachtberger \textit{et al.} \cite{schlachtberger2018cost} evaluated the influence of weather data, cost parameters, and policy constraints to the cost-optimal highly renewable European electricity system. Although the mentioned sensitivities were observed, the authors found the total system cost to be robust to different years of weather data and moderate changes in the cost assumptions. 

In the fourth place, a coarse-grained network is implemented due to computational limits, where every country is aggregated into a single node. This neglects possible problems such as national transmission network bottlenecks and local differences in weather conditions. H{\"o}rsch \textit{et al.} \cite{horsch2017role} explored the role of spatial scale in joint optimisations of generation and transmission for highly renewable Europe. They showed that spatial resolution has little impact to the total system cost, but more pronounced impact to the wind/solar mix.

In the fifth place, the \co{} emissions produced over the lifetime of VRES generators are not included when computing the system emissions. Louwen \textit{et al.} \cite{louwen2016re} presented a review of PV development, analysing the energy demand and \co{} emissions associated with PV production. The current \co{} footprint for PV is around 20~g\co{}/KWh\el, which is 1/24 compared with burning gas in OCGT power plants (480~g\co{}/KWh\el, efficiency of 0.39). According to Wiser \textit{et al.} \cite{wiser2011wind}, the median value of \co{} emissions for wind is 12~g\co{}/KWh\el, which is 1/40 compared to gas. If the energy necessary to manufacture the VRES generators comes from a highly renewable energy system those values would be even lower. 

Finally, grid losses have been neglected in the model since PyPSA is programmed as a linear optimisation. Grid losses are typically small. For instance, the losses from a volume of 60 TWkm and 331 TWkm have been estimated to be 0.4\% and 2.1\% of the electricity annual demand, respectively \cite{Gils2017a}.

\section{Conclusions} \label{sec:conclusion}
A simplified hourly-resolved one-node-per-country network has been employed to investigate the optimal design of coupled electricity and heating system for Europe. The \WH layout is introduced, which guarantees that each country is renewable self-sufficient to a certain extent, while the best wind/solar mix design within each country is sought by the optimisation. 

It has been shown that in order to achieve ambitious \co{} reduction targets in Europe, installing large amounts of Variable Renewable Energy Sources (VRES) is not enough but a significantly high \co{} price is required to disincentivise the use of gas. The cost-optimal system configuration to attain 5\% \co{} emissions, relative to 1990, has been identified. This target configuration requires a \co{} price equal to 380~\euro/t\co{}. Since no biomass to produce electricity is included in the model, the \co{} price must be high enough to alter the merit order between an expensive supply of heat through power-to-heat technologies and a cheap use of gas for heating. For the 5\% target configuration (380~\euro/t\co{}), the gross VRES penetration is 0.64 and the optimum solar/wind mix is 0.8. More importantly, the flatness around cost-optimal \co{} price allows policy makers to choose a wide range of \co{} prices and VRES configurations without significantly affecting the total system costs. 
%Restricting transmission volumes to todays volume has a non-negligible impact to the system and removing PTH technologies leads to a system with high curtailment, which emphasises the key role of PTH technologies to decarbonise the heating sector. 

For the target configurations attaining 5\%, 10\%, and 20\% \co{} emissions relative to 1990, the spatial distribution of system cost and primary energy have been examined in detail. Most investments go into renewable installation and power-to-heat technologies, which are incentivised by strict \co{} emission limits. For 5\% \co{} emissions, the heating sector is supplied mostly by heat pumps, while resistive heaters and gas boiler are used mainly to cover the peak heating demand.

This study opens up several avenues for future extensions. As mentioned in Section \ref{subsec:configuration}, wind is more favoured as the system evolves towards highly renewable penetration. The mechanism behind this phenomenon needs to be further analysed. The heating sector is mostly powered by electricity generated from wind and solar. It is interesting to investigate how much and when the different forms of renewable generations contribute to this conversion. It has already been used in \cite{Tranberg2018} to study the storage usage in a low-carbon electricity system. One could also include weather data altered by climate change \cite{schlott2018impact, kozarcanin2018climate}, or consider possible heat saving scenarios to discuss the impact of alternative heating demand on the optimal highly renewable system configuration. Although the target configurations for different \co{} reductions have been identified, the time periods to achieve those goals and transition pathways remain unclear, thus needing to be addressed.

\section*{Acknowledgements}
K. Z., M. Victoria, T. Brown, M. Greiner and G. B. Andersen are fully or partially funded by the RE-INVEST project, which is supported by the Innovation Fund Denmark under grant number 6154-00022B. T.B. also acknowledges funding from the Helmholtz Association under grant no. VH-NG-1352. The responsibility for the contents lies solely with the authors. 

\section*{References}
%\bibliography{elsarticle-template.bbl}
\bibliography{mybibfile}

\end{document}